\documentstyle[12pt]{article}

\makeatletter

\expandafter\ifx\csname mathrm\endcsname\relax\def\mathrm#1{{\rm #1}}\fi

\makeatother

\def\beq{\begin{equation}}
\def\eeq{\end{equation}}
\def\beqar{\begin{eqnarray}}
\def\eeqar{\end{eqnarray}}
\def\barr#1{\begin{array}{#1}}
\def\earr{\end{array}}
\def\bfi{\begin{figure}}
\def\efi{\end{figure}}
\def\btab{\begin{table}}
\def\etab{\end{table}}
\def\bce{\begin{center}}
\def\ece{\end{center}}
\def\nn{\nonumber}
\def\disp{\displaystyle}
\def\text{\textstyle}

\def\al{\alpha}

\def\Ga{\Gamma}
\def\de{\delta}
\def\De{\Delta}

\def\la{\lambda}
\def\si{\sigma}
\def\Si{\Sigma}

\def\refeq#1{\mbox{(\ref{#1})}}
\def\reffi#1{\mbox{Fig.~\ref{#1}}}
\def\reffis#1{\mbox{Figs.~\ref{#1}}}

\def\refse#1{\mbox{Sect.~\ref{#1}}}

\def\refapp#1{\mbox{App.~\ref{#1}}}
\def\citere#1{\mbox{Ref.~\cite{#1}}}
\def\citeres#1{\mbox{Refs.~\cite{#1}}}

\renewcommand{\O}{{\cal O}}

\def\mathswitchr#1{\relax\ifmmode{\mathrm{#1}}\else$\mathrm{#1}$\fi}

\newcommand{\PW}{\mathswitchr W}

\newcommand{\PH}{\mathswitchr H}

\def\mathswitch#1{\relax\ifmmode#1\else$#1$\fi}

\newcommand{\MW}{\mathswitch {M_\PW}}

\newcommand{\MH}{\mathswitch {M_\PH}}

\newcommand{\phrd}[1]{Phys.\ Rev.\ {\bf D#1}}
\newcommand{\phr}[1]{Phys.\ Rev.\ {\bf #1}}
\newcommand{\phrl}[1]{Phys.\ Rev.\ Lett.\ {\bf #1}}
\newcommand{\nphb}[1]{Nucl.\ Phys.\ {\bf B#1}}
\newcommand{\nphbps}[1]{Nucl.\ Phys.\ {\bf B} (Proc.\ Suppl.) {\bf #1B}}

\newcommand{\phlb}[1]{Phys.\ Lett.\ {\bf B#1}}

\newcommand{\zphc}[1]{Z.\ Phys.\ {\bf C#1}}
\newcommand{\aph}[1]{Ann.\ Phys.\ (NY) {\bf #1}}

\newcommand{\phrp}[1]{Phys.\ Rep.\ {\bf #1}}
\newcommand{\ptph}[1]{Prog.\ Theor.\ Phys.\ {\bf #1}}
\newcommand{\fp}[1]{Fortschr.\ Phys.\ {\bf #1}}

\def\tr#1{\,\mbox{tr}\left\{#1\right\}}
\def\Tr#1{\,\mbox{Tr}\left\{#1\right\}}
\newcommand{\spc}{\phantom{{}+{}}}
\newcommand{\dfun}{{\cal D}}
\newcommand{\lag}{{\cal {L}}}
\newcommand{\leff}{\lag_{\rm eff}}
\newcommand{\lreff}{\lag_{\rm eff}^{\rm ren}}
\newcommand{\lcth}{\lag_{\rm\hH}^{\rm ct}}
\newcommand{\Det}{\,\mbox{Det}\,}
\newcommand{\DW}{{\hat{D}_W}}
\newcommand{\Dp}{{\hat{D}_\varphi}}
\newcommand{\vp}{\varphi}
\newcommand{\hc}{\hat{\chi}}
\newcommand{\hvp}{\hat{\varphi}}
\newcommand{\hH}{\hat{H}}
\newcommand{\hW}{\hat{W}}
\newcommand{\hU}{\hat{U}}
\newcommand{\hV}{\hat{V}}
\renewcommand{\ss}{\scriptscriptstyle}
\newcommand{\intdp}{\int\frac{d^4 p}{(2\pi)^4}\,}
\renewcommand{\c}{\zeta}
\newcommand{\ord}[1]{\O(\c^{#1})}
\newcommand{\OH}[1]{\O(\MH^{#1})}
\newcommand{\dpp}{(x,\partial_x+i p)}
\newcommand{\tth}{ \tilde{\tilde{\Delta}}_H\dpp}
\newcommand{\nl}{\nn\\&&{}}
\newcommand{\tde}{\De_{\MH}}
\newcommand{\fac}{\frac{1}{16\pi^2}}

\def\Re{\mathop{\mathrm{Re}}\nolimits}

\def\draftdate{\relax}
\def\mda{\relax}
\def\mua{\relax}
\def\mla{\relax}
\def\draft{
\def\thtystars{******************************}
\def\sixtystars{\thtystars\thtystars}
\typeout{}
\typeout{\sixtystars**}
\typeout{* Draft mode!
         For final version remove \protect\draft\space in source file *}
\typeout{\sixtystars**}
\typeout{}
\def\draftdate{\today}
\def\mua{\marginpar[\boldmath\hfil$\uparrow$]%
                   {\boldmath$\uparrow$\hfil}%
                    \typeout{marginpar: $\uparrow$}\ignorespaces}
\def\mda{\marginpar[\boldmath\hfil$\downarrow$]%
                   {\boldmath$\downarrow$\hfil}%
                    \typeout{marginpar: $\downarrow$}\ignorespaces}
\def\mla{\marginpar[\boldmath\hfil$\rightarrow$]%
                   {\boldmath$\leftarrow $\hfil}%
                    \typeout{marginpar: $\leftrightarrow$}\ignorespaces}
\overfullrule 5pt
\oddsidemargin -15mm
\marginparwidth 29mm
}

\def\stars{\strut\leaders\hbox{*}\hfill\strut}
\def\starline{\hfil\strut\hfil\hbox to \textwidth {\stars}\hfil}

\renewcommand{\theequation}{\thesection.\arabic{equation}}
\newcounter{saveeqn}

\newcommand{\eqnew}{\setcounter{equation}{0}}

\makeatletter

\def\eqnarray{\stepcounter{equation}\let\@currentlabel=\theequation
\global\@eqnswtrue
\global\@eqcnt\z@\tabskip\@centering\let\\=\@eqncr
$$\halign to \displaywidth\bgroup\hskip\@centering
  $\displaystyle\tabskip\z@{##}$\@eqnsel&\global\@eqcnt\@ne
  \hskip 2\arraycolsep \hfil${##}$\hfil
  &\global\@eqcnt\tw@ \hskip 2\arraycolsep $\displaystyle\tabskip\z@{##}$\hfil
   \tabskip\@centering&\llap{##}\tabskip\z@\cr}
\def\appendix{\par
 \setcounter{section}{0} \setcounter{subsection}{0}
 \def\thesection{\Alph{section}}}

\makeatother

\begin{document}

\begin{titlepage}
\title{Deriving Non-decoupling Effects of Heavy Fields\\from
the Path Integral:\\
a Heavy Higgs Field in an SU(2) Gauge Theory}
\author{Stefan Dittmaier${}^1$\thanks{
E-Mail: didi@physik.uni-wuerzburg.de}
{}~and Carsten Grosse-Knetter${}^2$\thanks{
E-Mail: knetter@lps.umontreal.ca}~\thanks{
On leave of absence from the Universit\"at Bielefeld}
\\[5mm]
\normalsize ${}^1$Universit\"at Bielefeld, Fakult\"at f\"ur Physik,\\
\normalsize Postfach 10 01 31, D-33501 Bielefeld, Germany\\[5mm]
\normalsize ${}^2$Universit\'e de Montr\'eal, Laboratoire de Physique
Nucl\'eaire,\\
\normalsize C.P. 6128, Montr\'eal, Qu\'ebec, H3C 3J7, Canada}
\date{BI-TP 95/01\\UdeM-GPP-TH-95-15\\
hep-ph/9501285\\January 1995}
\maketitle
\thispagestyle{empty}

\begin{abstract}
We describe a method to remove
non-decoupling heavy fields from a quantized field
theory and to construct a low-energy one-loop effective Lagrangian
by integrating out the
heavy degrees of freedom in the path integral. We apply this method
to the Higgs boson in a spontaneously
broken SU(2) gauge theory (gauged linear $\si$-model).
In this context, the background-field method is generalized to the
non-linear representation of the Higgs sector
by applying (a generalization of) the Stueckelberg formalism.
The (background) gauge-invariant renormalization is discussed.
At one loop the $\log\MH$-terms of the heavy-Higgs limit of this model
coincide with the UV-divergent terms of
the corresponding gauged {\em non-linear\/} $\si$-model, but vertex
functions differ in addition by finite (constant) terms in both models.
These terms are also derived by our method.
Diagrammatic calculations of some vertex functions are presented as
consistency check.
\end{abstract}
\end{titlepage}

\section{Introduction}

Effective Lagrangians are used in order to describe the low-energy
effects in a theory with heavy particles. The
effective Lagrangian contains only light particles and
is an approximation to an underlying theory at
energy scales much
lower than the heavy particles' masses. In general
the complete theory is not known and thus the effective Lagrangian
contains undetermined parameters.
On the other hand, if one knows the underlying
theory, the free parameters can be calculated.

There are two different possibilities to construct a low-energy
effective Lagrangian of a given theory:
\begin{itemize}
\item One may integrate out the heavy particles by diagrammatic
methods, i.e. one calculates the contributions of all
{}Feynman diagrams
with internal heavy particles to
Green functions at a given
loop order (usually at one loop) and finds the parameters of the
effective theory by matching it to the full theory \cite{hemo,bisa}.
\item A more fundamental approach is to use functional methods,
i.e.\ to integrate out the heavy particles in the path integral. This
generates a functional determinant. The contributions of this
determinant to the effective Lagrangian can then be expanded in
(inverse) powers of the heavy particles' masses
\cite{bisa,gale,fume,chan,chey}.
\end{itemize}
In the present article we focus on the functional approach. We
describe a general and simple
method to integrate out
non-decoupling heavy fields in the path
integral and to obtain a
one-loop effective Lagrangian.
We explain this method by applying it to
the Higgs boson in a spontaneously broken
SU(2) gauge theory, assuming that this
boson
is very heavy.
A large number of articles about functional methods exist, e.g.\
\citeres{bisa,gale,fume,chan,chey}. We
partially
make use of methods developed in some of these
works, in particular of those in
\citere{chan} and \citere{chey}, however, as a whole and in its full detail
our procedure has not been described before.

In order to integrate out the heavy fields we use the background-field
method (BFM) \cite{bfm1,bfm2,bfm3,bfm4,bfm5}
in which the fields are split into a (classical)
background part, which corresponds to
tree lines in Feynman graphs, and a quantum
part, which corresponds to lines
inside loops. Thus, one
has to consider only
the part in the Lagrangian which is
quadratic in the quantum fields in order to construct vertex functions
at the one-loop level. The quantum fields associated
with the heavy particles can be integrated out by Gaussian
integration.
The resulting effective Lagrangian still contains the heavy
background fields, which can
be easily eliminated either diagrammatically by
a propagator expansion of the corresponding tree-lines or equivalently by
applying the classical equations
of motion for the background fields in lowest order.

In most of the existing works not only the heavy quantum fields
but all fields are integrated out.
Although also in this case an
$1/M$-expansion can formally be done, it
will not really be appropriate if not all
masses are large. However, an $1/M$-expansion is a useful tool if $M$
is a heavy particle's mass. Actually, some care has to be taken when
integrating out only a part of the quantum fields, viz.\ one has to
diagonalize the Lagrangian, such that the terms which contain
both heavy and light quantum fields
are
removed. This goal can be achieved
by appropriately shifting the quantum fields
in the path integral
\cite{gale,chey}.

A phenomenologically very important field of
application for this procedure
is the electroweak standard model (SM) provided that the Higgs boson
has a
mass much larger than the gauge bosons.
In this case the Higgs boson can be integrated out
and the corresponding low-energy Lagrangian can be constructed. This
will be done in a forthcoming paper \cite{smpaper}. In this article
we consider a toy model, namely the SU(2) gauged linear
$\si$-model (GLSM), which is similar to the SM, but simpler because
of the missing U(1) part and the corresponding mixing in the neutral
sector. We integrate out the Higgs boson in this model at the
one-loop level and determine
the effective Lagrangian to the order $\MH^0$ (which includes
$\log\MH$), i.e.\ we calculate those terms that contribute in the
limit $\MH\to\infty$.
The discussion of this toy model has the advantage,
that the physically important and interesting features
are essentially the same as in the SM, but the calculations are
technically less involved. Therefore,
this model is well suited for
explaining our method in full detail.

It is well known that the limit of the gauged linear
$\si$-model for $\MH\to\infty$ at
tree-level is the gauged {\em non-linear\/}
$\sigma$-model (GNLSM) \cite{bash,apbe},
which is formally constructed by
disregarding the Higgs boson in the non-linear
parametrization of the GLSM.
At the one-loop level,
in \citere{apbe} the assumption has been made
that the logarithmically
divergent contributions to S-matrix elements in the
non-renormalizable
GNLSM correspond to the logarithmically $\MH$-dependent contributions
in the GLSM, provided that
the poles in $(D-4)$ -- with $D$ being the space-time dimension in
dimensional regularization -- are appropriately replaced by $\log\MH^2$.
We find that this is indeed the case, however that
there are additional finite and $\MH$-independent differences between the
GNLSM and the GLSM at one loop. Thus, the GNLSM with the replacement
$\mbox{$2/(4-D)$}-\gamma_E+\log(4\pi)\to \log(\MH^2/\mu^2)$
is {\em not}
identical to the heavy-Higgs limit of the GLSM beyond tree-level.

This result has recently been derived in \citere{hemo} for the SM
by diagrammatical calculations.
Here, we derive it for the SU(2) model more directly by functional methods.
Comparing both methods we find that the functional method has many
advantages. E.g.\ all calculations can be done within the convenient
matrix notation, the
$1/\MH$-expansion becomes very straightforward,
and the use of the BFM enables us to chose the unitary gauge for the
background fields by applying the Stueckelberg formalism
\cite{stue1,stue2}, which removes the background Goldstone fields from
intermediate calculations.
Inverting the Stueckelberg transformation at the end, we recover the
background Goldstone fields and obtain the result for an arbitrary
background gauge.

This article is organized as follows: In \refse{bfmstf} we describe the
background-field method and the Stueckelberg formalism for the SU(2)
gauged linear $\sigma$-model and determine the part of the
Lagrangian, which contributes to one-loop
amplitudes. In \refse{diag} we diagonalize this Lagrangian, i.e.\
we remove all terms containing both light and heavy quantum fields.
In \refse{helim} we integrate out the quantum Higgs field and
construct the effective Lagrangian which parametrizes the one-loop
effects of the heavy Higgs boson. In \refse{invstue} this
Lagrangian is written in a manifestly gauge-invariant form.
In \refse{ren} we carry out the (gauge-invariant) renormalization.
In \refse{bghelim} the background Higgs field is
eliminated,
which yields the final
effective Lagrangian.
In \refse{examples}
we check the result of our functional procedure by comparing
it with diagrammatical calculations for some vertex functions.
\refse{dis} contains the discussion of the result
and \refse{sum} the summary. In \refapp{ints} the explicit form of
the Feynman integrals occurring in the calculations are given.

\section{The Background-Field Method and the Stueckelberg Formalism}
\label{bfmstf}

We consider the Lagrangian  of an SU(2) gauged linear $\sigma$-model
(GLSM) without fermions.
Using the matrix notation the Lagrangian is parametrized by
\beq
\lag= -\frac{1}{2} \tr{W_{\mu\nu}W^{\mu\nu}}
+\frac{1}{2}\tr{(D_\mu\Phi)^\dagger(D^\mu\Phi)}
+\frac{1}{2}\mu^2\tr{\Phi^\dagger\Phi}
-\frac{1}{16}\lambda\left(\tr{\Phi^\dagger\Phi}\right)^2,
\label{glsm}\eeq
with the gauge fields, field-strength tensors and covariant derivatives
given by
\beqar
W^\mu&=&\frac{1}{2}W^\mu_i \tau_i, \nn\\
W^{\mu\nu}&=&\partial^\mu W^\nu-\partial^\nu W^\mu-ig[W^\mu,W^\nu],\nn\\
D^\mu\Phi&=&(\partial^\mu-igW^\mu)\Phi,
\label{wpara}
\eeqar
where $\tau_i$ denote the Pauli matrices.
The scalar doublet $\Phi$ is linearly parametrized by
\beq
\Phi=\frac{1}{\sqrt{2}}\left((v+H){\bf 1}+2i\varphi\right),
\label{philin}
\eeq
with
\beq
\vp=\frac{1}{2}\varphi_i\tau_i,\qquad v=2\sqrt{\frac{\mu^2}{\lambda}},
\label{phipara}
\eeq
where $v$ is the non-vanishing vacuum expectation value,
$H$ is the (physical) Higgs field
and the $\varphi_i$ are the (unphysical) Goldstone fields. However, in
order to construct the heavy-Higgs limit of this model, it is more
useful to use instead the non-linear parametrization of the scalar
sector \cite{stue2}, which is specified by
\beq
\Phi=\frac{1}{\sqrt{2}}(v+H)U,
\label{phinl}
\eeq
where $H$ (unlike in the linear parametrization) is an SU(2) singlet
and the Goldstone fields are arranged
to form a
unitary matrix $U$. A convenient representation of $U$
is given by the exponential form
\beq
U=\exp\left(2i\frac{\varphi}{v}\right).
\label{U}
\eeq
The linear parametrization \refeq{philin} und the non-linear one
\refeq{phinl}, \refeq{U} are physically equivalent, i.e.\ they yield
the same S-matrix, although Green functions are different. The reason
for this is that the Jacobian determinant of the field transformation,
which relates \refeq{philin} to  \refeq{phinl}, only yields
contributions proportional to $\delta^4(0)$ \cite{stue2},
which vanish in dimensional regularization.
{}From
the GLSM with the parametrization \refeq{phinl}
the corresponding gauged {\em non-linear\/}
$\sigma$-model (GNLSM) \cite{bash,apbe}
can be obtained simply by disregarding the Higgs field
rendering the modulus of $\Phi$ constant,
$\tr{\Phi^\dagger\Phi}\to v^2=const.$
The GNLSM is the $\MH\to\infty$ limit of the GLSM
{\em at tree level\/} \cite{apbe}. It will turn out later in this
work that the $\MH\to\infty$ limit of the GLSM at one loop is the
GNLSM plus some effective interaction terms. Therefore, the non-linear
parametrization, \refeq{phinl} with \refeq{U},
is the more adequate one
for integrating out the Higgs boson. With this
parametrization the Lagrangian \refeq{glsm} becomes
\beqar
\lag&=& -\frac{1}{2} \tr{W_{\mu\nu}W^{\mu\nu}} + \frac{1}{2}
(\partial_\mu H)(\partial^\mu H) + \frac{1}{4}
(v+H)^2\tr{(D_\mu U)^\dagger(D^\mu U)} \nn\\
&&{}+\frac{1}{2}\mu^2 (v+H)^2
-\frac{1}{16}\lambda(v+H)^4,
\label{glsmnl}\eeqar
where the Goldstone fields $\vp_i$
occur only in the kinetic term
of the scalars owing to the unitarity of $U$.

The Lagrangian \refeq{glsmnl} contains terms cubic and quartic in the
Higgs field $H$. Thus, the integral which has to be performed when
integrating out $H$ in the path integral is not of Gaussian type.
At one loop order
this problem can be circumvented
by applying the background-field method (BFM)
\cite{bfm1,bfm2,bfm3,bfm4,bfm5}, in which the fields are split
into (classical) background fields and  quantum fields such that the
functional integration is only performed over the latter. In this
formalism the quantum fields appear only inside loops and the
background fields only
on tree lines. This means that terms higher
than quadratic in the quantum fields only contribute to higher loop
orders, but can be neglected in one-loop calculations. Therefore, the
Lagrangian to be considered is quadratic in the quantum Higgs field,
which can then be integrated out by Gaussian integration.

The BFM for the SM has been formulated in
\citeres{bfm3,bfm4,bfm5},
based on the linear parametrization of the
Higgs-Goldstone sector \refeq{philin}. For our
purpose we have to
modify this procedure and to adapt it to the non-linear parametrization
\refeq{phinl} with \refeq{U}.
As usual, we
split
the gauge and Higgs fields:
\beq
W^\mu \to \hat{W}^\mu + W^\mu,\qquad H \to \hat{H} + H,
\label{whsplit}\eeq
where $\hat{W^\mu}$ and $\hat{H}$ are the background fields and
$W^\mu$ and $H$ are the quantum fields. However,
for the Goldstone fields \refeq{U} it is more appropriate to do a
non-linear split, namely \cite{chey}
\beq
U \to \hat{U}U,
\label{Usplit}
\eeq
where $\hat{U}$ and $U$ are parametrized in terms of the
background and quantum Goldstone fields $\hat{\varphi}_i$ and
$\varphi_i$ according to
\refeq{phipara} and \refeq{U}, respectively.

The advantage of the non-linear split \refeq{Usplit} is the
following: we can now apply the Stueckelberg formalism
\cite{stue1,stue2} -- or stricly speaking a generalization of this
formalism to the BFM -- in order to remove the background Goldstone
fields $\hat{\varphi}_i$ from the Lagrangian.
We apply the following transformation
to the background and quantum vector fields \cite{chey}:
\beq
\hat{W}^\mu \to \hat{U}\hat{W}^\mu\hat{U}^\dagger +\frac{i}{g}
\hat{U}\partial^\mu\hat{U}^\dagger, \qquad
W^\mu \to \hat{U}W^\mu\hat{U}^\dagger.
\label{sttrafo}
\eeq
One can easily see that the covariant derivatives and field-strength
tensors, as defined in \refeq{wpara}, transform under this
Stueckelberg transformation as
\beq
D^\mu\hat{U}U \to \hat{U} D^\mu U, \qquad
(\hat{W}^{\mu\nu}+W^{\mu\nu})\to \hat{U}
(\hat{W}^{\mu\nu}+W^{\mu\nu}) \hat{U}^\dagger .
\label{sttrafo2}
\eeq
Note that the covariant derivative $D^\mu$ in (\ref{sttrafo2}) contains
the sum $\hW^\mu+W^\mu$ of the background and quantum gauge fields.
Thus, the effect of the Stueckelberg transformation \refeq{sttrafo}
is simply to remove
the background Goldstone fields from the Lagrangian,
i.e.\ $\hat{U}\to \bf 1$, while leaving
everything else
unaffected.

This means
that such a
Stueckelberg transformation corresponds to the choice of the unitary
gauge (U-gauge)
for the background fields. It is an advantage of the BFM that
different gauges can be chosen for the background fields and the
quantum fields \cite{bfm2,bfm4,bfm5}. While a choice of the U-gauge for
the quantum fields would complicate loop calculations, the U-gauge
for the background fields causes no problems because these fields do
not occur in loops. Instead,
the background U-gauge
reduces the number
of terms to be considered in the subsequent treatment (or,
equivalently, the number
of Feynman diagrams in a diagrammatic procedure). Actually, we will
choose a generalized $\rm R_\xi$-gauge for the quantum fields later.
By doing the
Stueckelberg transformation \refeq{sttrafo} inversely after all
calculations have been done,
the $\hat{\varphi}_i$ can easily be recovered and an
arbitrary other gauge for the background fields may be chosen.

The application of the Stueckelberg formalism within the BFM has
another important advantage: it automatically ensures the invariance
of the effective action under gauge transformations of the
background fields. In the conventional formalism, the Faddeev--Popov
quantization, i.e.\ the introduction of a gauge-fixing and a ghost
term destroys the gauge invariance of the effective action.
However, in the BFM it only destroys the invariance with respect to
gauge transformations of the quantum fields but a gauge-fixing term
for the quantum fields
can be chosen such that background gauge invariance is still
maintained
\cite{bfm2,bfm3,bfm4,bfm5}.
This invariance implies then simple Ward
identities \cite{bfm2,bfm4,bfm5}. In general, the demand of background
gauge invariance restricts the choice of the gauge-fixing term.
However, after applying the Stueckelberg transformation
\refeq{sttrafo} all quantities are automatically background
gauge-invariant. This can be seen as follows: apply an SU(2) gauge
transformation
to the background fields {\em
before\/} the Stueckelberg transformation
\refeq{sttrafo}
is done:
\beq
\hat{H} \to \hat{H}, \qquad
\hat{{U}} \to S \hat{{U}}, \qquad
\hat{{W}}^\mu \to S \hat{{W}}^\mu S^\dagger +\frac{i}{g}
S\partial^\mu S^\dagger
\label{bgtrafo}
\eeq
with
\beq
S = \exp\left(ig\theta\right),
\qquad\mbox{with}\quad \theta=\frac{1}{2}\theta_i\tau_i,
\eeq
where $\theta_i$ are the group parameters,
and
transform the quantum fields according to
\beq
{W}^\mu \to S {W}^\mu S^\dagger .
\label{qredef}
\eeq
\refeq{sttrafo} implies
that the fields obtained {\em after\/}
the Stueckelberg transformation are
singlets under \refeq{bgtrafo}, \refeq{qredef}. Thus, an arbitrary
gauge-fixing term written in terms of these automatically fulfills
the requirement of background gauge invariance.

Next, we have to determine that part of the Lagrangian
\refeq{glsmnl}, which is relevant for one-loop calculations; i.e.
the part quadratic in the quantum fields.
While the pure background
part describes the tree-level effects, the part linear in the quantum
fields
is irrelevant, and the terms with third or higher
powers of the quantum fields contribute only at higher loop orders
\cite{bfm1,bfm2}. Inserting \refeq{whsplit}, \refeq{Usplit}
and \refeq{sttrafo} into \refeq{glsmnl} and defining
\beqar
\Dp^\mu X &=& (\partial^\mu-ig\hW^\mu) X,\nn\\
\DW^\mu X &=& \partial^\mu X -i g [ \hW^\mu, X],
\label{covder}
\eeqar
one arrives at
\beqar
\lag^{\rm 1-loop}&=& \tr{ W_\mu\left( g^{\mu\nu} \DW^2
-\DW^\mu\DW^\nu +2ig \hat{W}^{\mu\nu}
+g^{\mu\nu}\frac{1}{4}g^2(v+\hat{H})^2 \right) W_\nu} \nn\\
&&{}-\tr{\vp\left(\frac{1}{v^2}\hat{D}_{\vp\mu}(v+\hH)^2 \Dp^\mu
+g^2\frac{1}{v^2}(v+\hH)^2
\hW_\mu \hW^\mu\right)\vp}\nn\\
&&{}-\frac{1}{2}H\left(\partial^2-\mu^2
+\frac{3}{4}\la(v+\hH)^2
-\frac{1}{2}g^2\tr{\hW_\mu \hW^\mu}\right)H\nn\\
&&{}-2 g\frac{1}{v}(v+\hH) H \tr{\hW_\mu\partial^\mu\vp}\nn\\
&&{}+g^2 (v+\hH) H \tr{\hW_\mu W^\mu}\nn\\
&&{}-g \frac{1}{v} (2v+\hH) \hH \tr{W_\mu \partial^\mu\vp}
-2ig^2v\tr{W_\mu\hW^\mu\vp}
+gv\tr{\vp\DW^\mu W_\mu} \nn\\
&&{}+\lag_{\rm gf}+\lag_{\rm ghost}.
\label{lag1}
\eeqar
Now, we have to fix the gauge of the quantum fields by introducing an
appropriate gauge-fixing term $\lag_{\rm gf}$. Similar to the
procedure in the linear
parametrization \cite{bfm3, bfm4,bfm5},
we choose this term such that the last
term in
\refeq{lag1}, which contains a $W\vp$-mixing, is cancelled.
The gauge-fixing term is
\beq
\lag_{\rm gf}= - \frac{1}{\xi} \tr{\left(\DW^\mu W_\mu
+\frac{1}{2}\xi gv\vp\right)^2}.
\label{gfterm}
\eeq
Recall that $\lag_{\rm gf}$ in (\ref{gfterm}) is written down in the
background U-gauge, i.e.\ for
$\hat{U}={\bf 1}$;
its full form for arbitrary
background gauges is obtained by inverting the Stueckelberg
transformation (\ref{sttrafo}).
As mentioned above, the Stueckelberg formalism ensures that this term
and the corresponding ghost term are invariant under gauge
transformations of the background fields.

The ghost Lagrangian $\lag_{\rm ghost}$, which corresponds to the
gauge-fixing term $\lag_{\rm gf}$
\refeq{gfterm}, can be easily
derived as usual. Note that the ghosts neither couple to $H$
nor to $\hH$ since the Higgs field, which is an SU(2) singlet,
does not occur in
the
gauge-fixing term. Therefore, $\lag_{\rm gf}$
and $\lag_{\rm ghost}$ are identical in the GLSM and the GNLSM.
Moreover, the ghost term obviously contains in one-loop order
(i.e.\ reduced to its part quadratic in the quantum fields)
no other quantum fields then ghosts and remains unaffected by
all our manipulations.

Inserting these terms into \refeq{lag1} and
expressing the parameters $v$, $\mu^2$ and $\lambda$ through the
gauge-boson and Higgs masses $\MW$ and $\MH$,
respectively,
\beq
v = \frac{2\MW}{g}, \qquad \mu^2 = \frac{1}{2}\MH^2, \qquad \lambda =
\frac{g^2\MH^2}{2\MW^2},
\label{vmula}
\eeq
one finally finds:
\beqar
\lag^{\rm 1-loop}&=& \tr{W_\mu \Delta^{\mu\nu}_W
W_\nu}-\tr{\vp\Delta_\vp \vp} -\frac{1}{2}H \Delta_H H \nn\\&&{}+
H \tr{X_{\ss H\vp} \vp} +H \tr{ X^\mu_{\ss H W}W_\mu} +
\tr{W_\mu X^\mu_{\ss W\vp}\vp}+\lag_{\rm ghost},
\label{lag2}
\eeqar
with
\beqar
\Delta_W^{\mu\nu}&=& g^{\mu\nu} \left(\DW^2 +
\MW^2\left(1+\frac{\hH}{v}\right)^2\right)
+\frac{1-\xi}{\xi} \DW^\mu \DW^\nu
+2ig W^{\mu\nu}
,\nn\\
\Delta_\vp&=&
\partial^\mu\left(1+\frac{\hH}{v}\right)^2\partial_\mu
+\xi\MW^2
-2ig\left(1+\frac{\hH}{v}\right)^2\hW^\mu\partial_\mu, \nn\\
\Delta_H&=&
\partial^2+\MH^2
+\frac{3}{2}\MH^2\frac{\hH}{v}\left(2+\frac{\hH}{v}\right)
-\frac{1}{2}g^2 \tr{\hW^\mu\hW_\mu},\nn\\
X_{\ss H \vp}&=&-2 g\left(1+\frac{\hH}{v}\right)\hW^\mu\partial_\mu,
\nn\\
X_{\ss H W}^\mu&=&2g\MW\left(1+\frac{\hH}{v}\right)\hW^\mu, \nn\\
X_{\ss W \vp}^\mu&=&
-2\MW\frac{\hH}{v}\left(2+\frac{\hH}{v}\right)\partial^\mu
-4 i g \MW \hW^\mu.
\label{deltas}
\eeqar

\section{Diagonalization of the one-loop Lagrangian}
\label{diag}
\eqnew

The Lagrangian \refeq{lag2} contains terms linear and quadratic in the
quantum Higgs field $H$. Therefore, removing $H$
by doing the integration over
this degree of freedom in the path integral
results simply in Gaussian integration. However,
the presence of terms linear in $H$
would yield effective terms in which
inverses of the operators
$\Delta_i$ \refeq{deltas}
act on the other quantum fields and
which cannot be evaluated easily.

Therefore, before integrating out $H$ we will rewrite the Lagrangian
\refeq{lag2} such that $H$ appears only quadratically,
and
thus, after
Gaussian integration the operators $\Delta_i^{-1}$ only
act on the background fields.
The terms linear in $H$ can be removed by shifts
(which yield unit Jacobian determinants)
in the quantum fields \cite{gale,chey} as follows:
{}First, we remove the $H\vp$-term by shifting the $\vp$ field. The
form of the Lagrangian \refeq{lag2} suggests a shift:
\beq
\vp\to\vp +\frac{1}{2} \hat\Delta_\vp^{-1} X_{\ss H\vp}^\dagger H.
\label{pshift}
\eeq
Note that on the r.h.s.\ of (\ref{pshift}) it is not allowed simply
to insert the inverse $\Delta_\vp^{-1}$ of $\Delta_\vp$ since the
l.h.s.\ is a linear combination of Pauli matrices while the r.h.s.\
would not be. On the other hand, we never need the full inverse
$\Delta_\vp^{-1}$
-- acting on  the space of self-adjoint $2\times 2$ matrices --
but only its restriction $\hat\Delta_\vp^{-1}$ to
the three-dimensional
linear subspace
spanned by the Pauli matrices $\tau_i$.
It turns out to be very useful to define the projection operator $P$,
\beq
P X= \frac{1}{2} \tau_i \tr{\tau_i X},
\label{P}\eeq
which maps the $\tau$-matrices onto themselves but the unit matrix to
zero.
In particular, $P$ acts as the identity on $\vp$ \refeq{phipara}
and on $W_\mu$ \refeq{wpara}.
Splitting $\Delta_\vp$
into a lowest-order part $\Delta_0$,
which is proportional to the unit matrix,
and a remainder $\Pi$,
\beq
\Delta_\vp = \Delta_0 + \Pi,
\eeq
we can explicitly write down $\hat\Delta_\vp^{-1}$ in the form of a
perturbative series,
\beqar
\hat\Delta_{\vp}^{-1} &=& \disp
\Delta_0^{-1}P\;\sum_{n=0}^{\infty}\;(-\Pi\Delta_0^{-1}P)^n \nn\\
&=& \Delta_0^{-1}P \;-\; \Delta_0^{-1}P\Pi\Delta_0^{-1}P \;+\;
\Delta_0^{-1}P\Pi\Delta_0^{-1}P\Pi\Delta_0^{-1}P \;+\; \cdots.
\label{Pcor}
\eeqar
With these definitions we obtain
\beq
(P\Delta_{\vp}P)\,\hat\Delta_{\vp}^{-1} =
\hat\Delta_{\vp}^{-1}\,(P\Delta_{\vp}P) = P.
\label{ddinv}
\eeq
Note that in the operator equations \refeq{Pcor}  and \refeq{ddinv}
$P$ (which commutes with $\Delta_0$ and its inverse) acts on the
whole expression right of it.
In the following all inverse operators with a hat are defined
analogously, i.e.\ they are restricted to the
linear subspace of
the $\tau$-matrices.

Shifting $\vp$ as given in
\refeq{pshift} with \refeq{Pcor} yields
\beqar
\lag^{\rm 1-loop}&=& \tr{W_\mu \Delta^{\mu\nu}_W
W_\nu}-\tr{\vp\Delta_\vp \vp}
-\frac{1}{2}H \tilde{\Delta}_H H \nn\\&&{}
+H \tr { \tilde{X}^\mu_{\ss H W}W_\mu}  +
\tr{W_\mu X^\mu_{\ss W\vp}\vp}+\lag_{\rm ghost},
\label{lag3}\eeqar
with
\beqar
\tilde{\Delta}_H&=&\Delta_H-\frac{1}{2}\tr{ X_{\ss H\vp}
\hat\Delta^{-1}_{\vp}X^\dagger_{\ss H\vp}},
\label{dhtilde}\\
\tilde{X}_{\ss HW}^\mu &=& {X}_{\ss HW}^\mu
+\frac{1}{2}X_{\ss H\vp}\hat\Delta^{-1}_{\vp}X_{\ss
W\vp}^{\mu\dagger}.
\label{xhwtilde}
\eeqar
If one would now shift the $W$-field in order to remove the
$HW$-term, the $W\vp$-term would generate a new
$H\vp$-term. Thus, we first shift
$\vp$ in order to remove that
term,
\beqar
\vp \to \vp +\frac{1}{2}\hat\Delta_\vp^{-1}
X_{W\vp}^{\mu\dagger} W_\mu,
\label{pshift2}
\eeqar
and $\lag^{\rm 1-loop}$ becomes
\beq
\lag^{\rm 1-loop}= \tr{W_\mu \tilde{\Delta}^{\mu\nu}_W
W_\nu}-\tr{\vp\Delta_\vp \vp}
-\frac{1}{2}H \tilde{\Delta}_H H
+H \tr { \tilde{X}^\mu_{\ss H W}W_\mu}
+\lag_{\rm ghost}
\label{lag4}\eeq
with
\beq
\tilde{\Delta}^{\mu\nu}_W = {\Delta}^{\mu\nu}_W +\frac{1}{4}
X^\mu_{\ss W\vp}\hat\Delta_\vp^{-1}
X^{\nu\dagger}_{\ss W\vp}.
\label{dwtilde}\eeq
Now, we can remove the $HW$-term by
\beq
W^\mu \to W^\mu -\frac{1}{2}
\hat{\tilde{\Delta}}^{-1\mu\nu}_W\tilde{X}_{{\ss HW},\nu}H
\label{wshift}
\eeq
which yields
\beq
\lag^{\rm 1-loop}= \tr{W_\mu \tilde{\Delta}^{\mu\nu}_W
W_\nu}-\tr{\vp\Delta_\vp \vp}
-\frac{1}{2}H \tilde{\tilde{\Delta}}_H H+
\lag_{\rm ghost}
\label{lag5}\eeq
with
\beq
\tilde{\tilde{\Delta}}_H = \tilde{\Delta}_H +
\frac{1}{2}\tr{\tilde{X}_{{\ss HW},\mu}
\hat{\tilde{\Delta}}^{-1\mu\nu}_W\tilde{X}_{{\ss HW},\nu}^\dagger}.
\label{tth}
\eeq
Now, all terms linear in $H$ are removed, however, some terms which do
not contain $H$, viz.\ the $WW$ and the $W\vp$-term are also changed.
In order to reconstruct these terms in their original form we do the
shift
\refeq{pshift2}
inversely and finally find
\beq
\lag^{\rm 1-loop}= \tr{W_\mu \Delta^{\mu\nu}_W
W_\nu}-\tr{\vp\Delta_\vp \vp} -\frac{1}{2}H
\tilde{\tilde{\Delta}}_H H +
\tr{W_\mu X^\mu_{\ss W\vp}\vp}+\lag_{\rm ghost}.
\label{lagfin}\eeq

\section{Elimination of the Quantum Higgs field and
$\bf 1/\MH$-Expansion}
\label{helim}
\eqnew

Now, we carry out the functional integration over the field $H$ in the
path integral, which yields
\beqar
Z&\propto &\int\dfun H\, \dfun W_i^\mu\, \dfun \vp_i\, \exp\left[i\int d^4x
\,\lag^{\rm 1-loop}\right]\nn\\
&\propto &\int \dfun W_i^\mu\, \dfun \vp_i\, \left\{\Det
\left( i \tilde{\tilde{\Delta}}_H
\delta^{(4)}(x-y)\right)\right\}^{-\frac{1}{2}}
\exp\left[i\int d^4x
\lag^{\rm 1-loop}\big|_{H=0}\right]
\nn\\[.2em]
&\propto&
\int \dfun W_i^\mu\, \dfun \vp_i\;
\exp\left(iS_{\rm eff}\right)\;
\exp\left[
i\int d^4x
\lag^{\rm 1-loop}\big|_{H=0}
\right],
\label{pi2}
\eeqar
with
\beq
S_{\rm eff}= \frac{i}{2} \Tr {\log
\left(\tilde{\tilde{\Delta}}_H \delta^{(4)}(x-y)\right)},
\label{seff}
\eeq
where ``Tr'' denotes the functional trace, in distinction from
the genuine SU(2) trace ``tr''.

The functional trace and logarithm now have to be evaluated. Many
different attempts to
perform this
evaluation
exist in the literature
\cite{bisa,gale,fume,chan,chey}. Our procedure is essentially based
on the method developed in \citere{chan}. We write
\beqar
\tilde{\tilde{\Delta}}_H(x,\partial_x) \delta^{(4)}(x-y)
&=&\intdp \tilde{\tilde{\Delta}}_H(x,\partial_x)
\exp\{ip(x-y)\}\nn\\
&=&\intdp \exp\{ip(x-y)\}
\tilde{\tilde{\Delta}}_H\dpp.
\eeqar
Now, the trace can be determined:
\beq
\Tr {\log
\left(\tilde{\tilde{\Delta}}_H \delta^{(4)}(x-y)\right)}
= \int d^4 x \intdp  \log
\left(\tilde{\tilde{\Delta}}_H\dpp\right),
\eeq
and thus \refeq{seff} implies
\beq
\leff = \frac{i}{2}
\intdp \log \left(\tilde{\tilde{\Delta}}_H\dpp\right).
\label{leff}
\eeq
$\tilde{\tilde{\Delta}}_H\dpp$ can be expanded in terms of derivatives,
\beq
\tilde{\tilde{\Delta}}_H\dpp =
\sum_{n=0}^\infty \frac{(-i)^n}{n!}
\left[\frac{\partial^n}
{\displaystyle\partial{p_{\mu_1}}\ldots\partial{p_{\mu_n}}}
\tilde{\tilde{\Delta}}_H(x,ip)\right]
\partial_{\mu_1}\ldots\partial_{\mu_n},
\label{dexpand}\eeq
which will yield an expression like
\beq
\tilde{\tilde{\Delta}}_H\dpp = p^2-\MH^2
+\Pi(x,p,\partial_x).
\eeq
Since $\Pi$ commutes with $p^2-\MH^2$, the logarithm in \refeq{leff}
can now easily be expanded (i.e.\ without having to apply the
Baker-Haussdorf formula).
Dropping
an irrelevant
constant, one finds
\beq
\log \tilde{\tilde{\Delta}}_H\dpp
=\sum_{n=1}^\infty
\frac{(-1)^{n+1}}{n}\left(\frac{\Pi}{p^2-\MH^2}\right)^n.
\label{logexpand}
\eeq

Since we are interested in the heavy-Higgs limit of the GLSM, we only
need to consider those terms, which do not vanish for $\MH\to\infty$,
i.e.\ all terms that are of the order $\MH^0$ (which includes
$\log\MH$) while inverse powers of $\MH$ can be neglected. The
procedure described above yields integrals of the type
\beq
\intdp\frac{\displaystyle
p_{\mu_1}\ldots p_{\mu_{2k}}}{(p^2-\MH^2)^l(p^2-M^2)^m},
\qquad\mbox{with}\quad l >0 \quad \mbox{and} \quad
M^2=\MW^2\quad\mbox{or}\quad \xi\MW^2,
\label{int}
\eeq
which are explicitly given in \refapp{ints}.
The factors $(p^2-\MW^2)^{-1}$ stem from the expansion of the
inverse operators $\hat{\Delta}_\vp^{-1}\dpp$
and $\hat{\Delta}_W^{-1\mu\nu}\dpp$ in $\tth$ according to
\refeq{dexpand} and the factors  $(p^2-\MH^2)^{-1}$ from the
expansion of the logarithm \refeq{logexpand}.
The integrals \refeq{int} are $\O(\MH^n)$ with
\beq
n=4+2(k-l-m)
\label{cond}
\eeq
if $n\ge0$, and $\OH{-2}$ or less if $n<0$
(see \refeq{vacint}).
This means, when expanding $\log \tilde{\tilde{\Delta}}_H
\dpp$, we only have to consider terms of
$\O(p^{-4})$ or higher powers of $p$.
{}Furthermore, some of the generated
terms contain the background Higgs field $\hH$. This will be
eliminated later
in \refse{bghelim} by the replacement
\beq
\hH\,\to\,\frac{g\MW}{\MH^2}\tr{\hW_\mu\hW^\mu}+\O(\MH^{-4}).
\label{hm2}
\eeq
This means, each $\hH$ contributes two negative powers of
$\MH$. Finally, there is an explicit $\MH$-dependence arising from
the couplings of the background Higgs to the quantum Higgs.
Therefore, when expanding $\log \tilde{\tilde{\Delta}}_H
\dpp$ we determine the leading powers
of $p$, $\hH$ and $\MH$ for each term
generated and introduce  an auxiliary parameter
$\c$,
which counts these powers according to
\beq
p_\mu \to \c,\qquad \MH \to \c, \qquad \hH \to \c^{-2}.
\label{count}
\eeq
Then, we only have to consider contributions up to $\O(\c^{-4})$
and can
neglect higher negative powers of $\c$. In particular this
means, that $\tilde{\tilde{\Delta}}_H
\dpp$ only has to be expanded up to
$\O(\c^{-2})$, because each contribution is multiplied with at
least one power of $(p^2-\MH^2)^{-1}$
in \refeq{logexpand}.

Using \refeq{deltas}, $\tilde{X}^\mu_{\ss HW}$ and
$\tilde{\Delta}_W^{\mu\nu}$, defined in \refeq{xhwtilde} and
\refeq{dwtilde}, can be expanded as
\beqar
\tilde{X}^\mu_{\ss HW}\dpp &=&{X}^\mu_{\ss HW}\dpp +\ord{-1},\\
\qquad \tilde{\Delta}_W^{\mu\nu}\dpp &=&{\Delta}_W^{\mu\nu}\dpp
+\ord{-2},\\
\Rightarrow  \qquad
\hat{\tilde{\Delta}}_W^{-1\mu\nu}\dpp &=&\hat{\Delta}_W^{-1\mu\nu}
\dpp+\ord{-6}.
\eeqar
Thus, $\tth$ given by \refeq{dhtilde} and \refeq{tth},
can be written as
\beqar \!\!\!\!\!\!
&&\tth =\nn\\ \!\!\!\!\!\!
&&\left(\Delta_H +
\frac{1}{2}\tr{{X}_{{\ss HW},\mu}
 \hat{\Delta}^{-1\mu\nu}_W{X}_{{\ss HW},\nu}}
-\frac{1}{2}\tr{X_{\ss H\vp}
\hat{\Delta}^{-1}_{\vp}X_{\ss H\vp}^\dagger}\right)\dpp+\ord{-3}.
\hspace{2em}
\label{tth2}
\eeqar
Now, we expand the quantities occurring in \refeq{tth2} as far as
necessary. The first term is simply
\beq
\Delta_H\dpp=-(p^2-\MH^2) +2ip_\mu\partial^\mu+
\partial^2+\frac{3g\MH^2}{2\MW}\hH
+\frac{3g^2\MH^2}{8\MW^2}\hH^2-\frac{g^2}{2} \tr{\hW^\mu\hW_\mu}.
\label{term1}
\eeq
Also, the second term can be expanded very easily.
Since $X_{\ss HW}\dpp$ is of
$\ord{0}$, we only need the
leading terms of  $X_{\ss HW}\dpp$ and $\hat{\Delta}_W^{-1}\dpp$ in
\refeq{tth2}.
Eq.~\refeq{deltas} yields
\beqar
X_{\ss HW}^\mu\dpp&=& 2 g \MW \hW^\mu +\ord{-2},\\
\Delta_W^{\mu\nu}\dpp&=&- g^{\mu\nu}(p^2-\MW^2)+\frac{\xi-1}{\xi}
p^\mu p^\nu +\ord{1},\\
\Rightarrow
\hat{\Delta}_W^{-1\mu\nu}\dpp&=& -\frac{g^{\mu\nu}}{p^2-\MW^2}P+
(1-\xi)\frac{p^\mu p^\nu}{(p^2-\MW^2)(p^2-\xi\MW^2)}P +\ord{-3}
\nn\\
&=& -\frac{g^{\mu\nu}}{p^2-\MW^2}P+
\frac{p^\mu p^\nu}{\MW^2}\left[
\frac{1}{p^2-\MW^2}-\frac{1}{p^2-\xi\MW^2}\right]P +\ord{-3}.
\hspace{3em}
\label{deltawinv}
\eeqar
Thus, the second term in \refeq{tth2} is
\beqar
\frac{1}{2}\tr{{X}_{{\ss HW}\mu}
\hat{\Delta}^{-1\mu\nu}_W{X}_{{\ss HW}\nu}}\dpp&=&
{}-2g^2\MW^2\frac{1}{p^2-\MW^2}\tr{\hW_\mu\hW^\mu}\nn\\&&{}
+2g^2\left[
\frac{p_\mu p_\nu}{p^2-\MW^2}-
\frac{p_\mu p_\nu}{p^2-\xi\MW^2}
\right]\tr{\hW^\mu\hW^\nu}\nn\\
&&{}+\ord{-3}.\label{term2}
\eeqar
The operator $P$
(\ref{P}) does not affect the result (\ref{term2})
in this order.
The expansion of the expressions occurring in the third term in
\refeq{tth2} is a bit more involved, because the leading term of
$X_{\ss H\vp}\dpp$ is $\ord{1}$ and thus also some
non-leading parts of $X_{\ss H\vp}\dpp$ and
$\hat{\Delta}^{-1}_\vp\dpp$ have to be taken into account:
\beqar
X_{\ss H\vp}\dpp&=&{}-2ig\hW_\mu p^\mu -2g\hW_\mu\partial^\mu
-i\frac{g^2}{\MW}\hH\hW_\mu p^\mu +\ord{-2},\label{xhp}\\
\Delta_\vp\dpp&=&{}-(p^2-\xi\MW^2) +2ip_\mu\partial^\mu +\partial^2\nn\\
&&{}+2g\hW_\mu p^\mu-2ig\hW_\mu\partial^\mu
-\frac{g}{\MW}\hH p^2 +\ord{-1},\\
\Rightarrow
\hat{\Delta}_\vp^{-1}\dpp &=&
{}-\frac{1}{p^2-\xi\MW^2}P
\nn\\&&{}
-\frac{1}{(p^2-\xi\MW^2)^2}P\!\left[
2ip_\mu\partial^\mu +\partial^2
+2g\hW_\mu p^\mu-2ig\hW_\mu\partial^\mu
-\frac{g}{\MW}\hH p^2 \right]\!P
\nn\\&&{}
-\frac{4p_\mu p_\nu}
{(p^2-\xi\MW^2)^3}P\!
\left[ -\partial^\mu \partial^\nu +i g
(\partial^\mu \hW^\nu +\hW^\mu \partial^\nu)
+g^2 \hW^\mu P \hW^\nu \right]\!P
\nn\\&&{}+\ord{-5}.
\label{deltapinv}
\eeqar
Since $P$ \refeq{P} acts as the identity as long as
there is exactly one Pauli matrix left or right of
it
and since $X_{\ss H\vp}\dpp$ is a linear combination of the
$\tau_i$, the $P$ in
\refeq{deltapinv} can be dropped everywhere
except for the very last term contributing to
$\hat{\Delta}_\vp ^{-1}\dpp$,
where two Pauli Matrices occur.
The third term in \refeq{tth2} becomes then
\beqar
\hspace{4em} && \hspace{-4em}
-\frac{1}{2}\tr{X_{\ss H\vp}
\hat{\Delta}^{-1}_{\vp}X_{\ss H\vp}^\dagger}\dpp=
\nn\\&& \spc
\frac{1}{p^2-\xi\MW^2}2g^2 \left[ \tr{(\partial_\mu\hW^\mu)^2}
+p_\mu p_\nu \tr{\hW^\mu\hW^\nu} \right]
\nn\\&&{}
-\frac{p_\mu p_\nu}{(p^2-\xi\MW^2)^2}
2 g^2 \left[
\tr{(\partial_\rho \hW^\mu)(\partial^\rho \hW^\nu)}
+4 \tr{(\partial^\mu\hW^\nu)(\partial_\rho
\hW^\rho)} \right.
\nn\\&&\qquad\qquad\qquad\qquad\qquad{} \left.
-2 i g
\tr{(\partial_\rho\hW^\nu)\hW^\rho\hW^\mu}\right]
\nn\\&&{}+\frac{p_\mu p_\nu p_\rho p_\sigma}{(p^2-\xi\MW^2)^3}
{8 g^2 \tr{(\partial^\mu
\hW^\nu)(\partial^\rho\hW^\sigma)}}
\nn\\&&{}+\ord{-3}.
\label{term3}
\eeqar

In determining \refeq{term3}, we have
already carried out some simplifications, which are strictly speaking
only justified in the full expansion of (\ref{leff}). More precisely,
we have dropped total derivative terms of $\ord{-2}$, which contribute
only to the linear term ($n=1$) in the expansion of the logarithm
\refeq{logexpand}. Such terms yield total derivatives of $\leff$ either.
{}Furthermore, we made use of the fact
that in all expressions the Lorentz indices can be arbitrarily
exchanged. In particular, this and the definition of $P$ \refeq{P}
imply that the only contribution
with this operator, which has the form
$p_\mu p_\nu p_\rho p_\sigma \tr{\hW^\mu \hW^\nu P \hW^\rho
\hW^\sigma}$, vanishes. Finally, the contributions from the
$\hH$ terms in $X_{\ss H\vp}\dpp$ \refeq{xhp} and in
$\hat{\Delta}^{-1}_{\vp}\dpp$
\refeq{deltapinv} cancel at this order,
because ${p_\mu p_\nu p^2} {(p^2-\MW^2)^{-2}}=
{p_\mu p_\nu} {(p^2-\MW^2)^{-1}} +\ord{-2}$.
Summing \refeq{term1}, \refeq{term2} and
\refeq{term3} we find the expansion for \refeq{tth2}.

Next, we expand $\log \tilde{\tilde{\Delta}}_H\dpp$ as in
\refeq{logexpand} and integrate over $p$ in order to obtain the
effective Lagrangian \refeq{leff}.
Only the first two terms in the expansion of
$\log \tilde{\tilde{\Delta}}_H\dpp$ yield $\ord{-4}$ (or higher)
contributions. Furthermore, only the $\hH$-term
in \refeq{term1},
the $\tr{\hW_\mu\hW^\mu}$-term in \refeq{term1}, and the $p_\mu p_\nu
\tr{\hW^\mu \hW^\nu}$-term in \refeq{term3} have to be considered in
the quadratic part of the expansion of the logarithm.
The integrals which have to be calculated are symmetric with respect
to exchange of Lorentz indices.
We use the notation
\beq
I_{klm}(\xi) g_{\mu_1\ldots\mu_{2k}}= \frac{(2\pi\mu)^{4-D}}{i\pi^2}
\int {d^D p}
\frac{\displaystyle
p_{\mu_1}\ldots p_{\mu_{2k}}}{(p^2-\MH^2)^l(p^2-\xi\MW^2)^m},
\label{intnot}
\eeq
with $g_{\mu_1\ldots\mu_{2k}}$ being the totally symmetric tensor
built of $g_{\mu\nu}$'s with
rank $2k$.
As indicated in (\ref{intnot}), we use dimensional regularization in
order to regularize occurring UV divergencies. $D$ denotes the space-time
dimension, and $\mu$ an arbitrary reference mass scale. The explicit
expressions for the occurring integrals are given in App.~\ref{ints}.
Dropping total derivatives,
contracting the Lorentz indices
and collecting all terms we find:%
\footnote{Note that the neglected remainder in \refeq{leff2}
is $\ord{-2}$ and not only $\ord{-1}$, since integrals like \refeq{intnot}
vanish for an odd number of $p_\mu$'s.}
\beqar
\leff&=&\spc\fac g\frac{3\MH^2}{4\MW^2}I_{010} \hH\nl
+\fac g^2\left(
\frac{3\MH^2 }{16\MW^2} I_{010}+\frac{9\MH^4}{16\MW^2}
I_{020}\right)\hH^2\nl
+\fac g^3
\left(-\frac{3\MH^2}{8\MW} I_{020} +\frac{3\MH^2}{2\MW}
I_{121}(\xi)\right)\hH\tr{\hW_\mu \hW^\mu}\nl
+\fac g^2\bigg(-\frac{1}{4}I_{010}
-\MW^2 I_{011}(1)+
I_{111}(1)\bigg)\tr{\hW_\mu \hW^\mu}\nl
+\fac g^2\bigg(- I_{112}(\xi) +4
I_{213}(\xi)\bigg)\tr{(\partial_\mu\hW_\nu)(\partial^\mu\hW^\nu)}
\nl
+\fac g^2 \bigg(I_{011}(\xi)-4 I_{112}(\xi)+8
I_{213}(\xi)\bigg)\tr{(\partial_\mu\hW^\mu)^2}\nl
+\fac  2ig^3 I_{112}(\xi) \tr{(\partial_\mu\hW_\nu)\hW^\mu\hW^\nu}
\nl
+ \fac 2 g^4I_{222}(\xi)\tr{\hW_\mu \hW_\nu}\tr{\hW^\mu \hW^\nu}\nl
+\fac g^4\left(\frac{1}{16}I_{020}-\frac{1}{2} I_{121}(\xi)+
I_{222}(\xi)\right)\left(\tr{\hW_\mu \hW^\mu}\right)^2\nl
+\ord{-2}.
\label{leff2}
\eeqar

\section{Inverting the Stueckelberg transformation}
\label{invstue}
\eqnew

Now,
we write these terms in a more convenient form by
introducing non-Abelian field-strength tensors \refeq{wpara}.
We first write
\beqar
\tr{(\partial_\mu\hW_\nu)(\partial^\mu\hW^\nu)}&=&
\frac{1}{2}\tr{\hW_{\mu\nu}\hW^{\mu\nu}}
+\tr{(\partial_\mu\hW^\mu)^2}\nn\\&&{}
+ig \tr{\hW_{\mu\nu}[\hW^\mu,\hW^\nu]}
-\frac{1}{2}g^2\tr{[\hW_\mu,\hW_\nu][\hW^\mu,\hW^\nu]},
\nn\\
\tr{(\partial_\mu\hW_\nu)\hW^\mu\hW^\nu}&=&\frac{1}{4}
\tr{\hW_{\mu\nu}[\hW^\mu,\hW^\nu]}+\frac{i}{4}g
\tr{[\hW_\mu,\hW_\nu][\hW^\mu,\hW^\nu]}.
\label{xx1}
\eeqar
The generated traces of four
Pauli matrices can be expressed through traces
of two
$\tau$'s by using the identity
\beq
\tr{\tau_i\tau_j\tau_k\tau_l}= \frac{1}{2}
\bigg(\tr{\tau_i\tau_j}\tr{\tau_k\tau_l}
-\tr{\tau_i\tau_k}\tr{\tau_j\tau_l}
+\tr{\tau_i\tau_l}\tr{\tau_j\tau_k}\bigg),
\eeq
which yields
\beqar
\tr{[\hW_\mu,\hW_\nu][\hW^\mu,\hW^\nu]}&=&2
\tr{\hW_\mu \hW_\nu \hW^\mu\hW^\nu}- 2\tr{\hW_\mu \hW^\mu
\hW_\nu\hW^\nu}\nn\\
&=& 2\tr{\hW_\mu\hW_\nu}\tr{\hW^\mu\hW^\nu}
-2\left(\tr{\hW_\mu\hW^\mu}\right)^2
\label{xx2}.
\eeqar
Inserting \refeq{xx1} and \refeq{xx2} into \refeq{leff2} we find
\beqar
\leff&=&\spc\fac g\frac{3\MH^2}{4\MW^2}I_{010} \hH\nl
+\fac g^2\left(
\frac{3\MH^2 }{16\MW^2} I_{010}+\frac{9\MH^4}{16\MW^2}
I_{020}\right)\hH^2\nl
+\fac g^3
\left(-\frac{3\MH^2}{8\MW} I_{020} +\frac{3\MH^2}{2\MW}
I_{121}(\xi)\right)\hH\tr{\hW_\mu \hW^\mu}\nl
+\fac g^2\bigg(-\frac{1}{4}I_{010}
-\MW^2 I_{011}(1)+
I_{111}(1)\bigg)\tr{\hW_\mu \hW^\mu}\nl
+\fac g^2\left(- \frac{1}{2}I_{112}(\xi) +2
I_{213}(\xi)\right)\tr{\hW_{\mu\nu}\hW^{\mu\nu}}
\nl
+\fac g^2 \bigg(I_{011}(\xi)-5 I_{112}(\xi)+12
I_{213}(\xi)\bigg)\tr{(\partial_\mu\hW^\mu)^2}\nl
+\fac i g^3 \left(-\frac{1}{2}I_{112}(\xi)+4I_{213}(\xi)\right)
 \tr{\hW_{\mu\nu}[\hW^\mu,\hW^\nu]}
\nl
+ \fac g^4\bigg(-4 I_{213}(\xi)+2I_{222}(\xi)\bigg)
\tr{\hW_\mu \hW_\nu}\tr{\hW^\mu \hW^\nu}\nl
+\fac g^4\bigg(\frac{1}{16}I_{020}-\frac{1}{2}I_{121}(\xi)
+4I_{213}(\xi)+
I_{222}(\xi)\bigg)\left(\tr{\hW_\mu \hW^\mu}\right)^2\nl
+\ord{-2}.
\label{leff4}
\eeqar

Then, we reintroduce the background Goldstone fields $\hat{\vp}_i$ by
inverting the
Stueckelberg transformation \refeq{sttrafo}, i.e.\ we transform the
background and quantum vector fields as
\beq
\hW^\mu\to \hat{U}^\dagger\hW^\mu\hat{U} +\frac{i}{g}
 \hat{U}^\dagger\partial^\mu\hat{U}=\frac{i}{g}
 \hat{U}^\dagger \Dp^\mu\hat{U},\qquad
W^\mu\to \hat{U}^\dagger W^\mu\hat{U},
\label{sttrafoinv}
\eeq
with $\Dp^\mu$ being defined in \refeq{covder}.
Obviously, the transformation of the quantum fields is only needed in
the remaining term $\lag^{\rm 1-loop}|_{H=0}$ in the path integral
\refeq{pi2}, while $\leff$ \refeq{leff4} only consists of background
fields. According to \citeres{hemo,apbe}, we introduce a shorthand
notation for the covariant derivative
\beq
\hat{V}^\mu=\left(\Dp^\mu \hat{U}\right)\hat{U}^\dagger.
\label{V}\eeq
The transformations of the vector fields, field-strength tensors
and derivatives in \refeq{leff4} under \refeq{sttrafoinv}
can be written as
\beqar
\hW^\mu&\to&\frac{i}{g} \hU^\dagger\hV^\mu\hU,\nn\\
\hW^{\mu\nu}&\to& \hU^\dagger\hW^{\mu\nu}\hU,\nn\\
\partial^\mu \hW_\mu &\to& \frac{i}{g}
\partial^\mu\left(\hU^\dagger\hV_\mu\hU\right)
=\frac{i}{g} \hU^\dagger \left(\DW^\mu\hV_\mu\right) \hU,
\label{divtrafos}\eeqar
with $\DW^\mu$ being defined in \refeq{covder}. Applying this to
\refeq{leff4}, we finally find the effective Lagrangian
\beqar
\leff&=&\spc\fac g\frac{3\MH^2}{4\MW^2}I_{010} \hH\nl
+\fac g^2\left(
\frac{3\MH^2 }{16\MW^2} I_{010}+\frac{9\MH^4}{16\MW^2}
I_{020}\right)\hH^2\nl
+\fac g
\left(\frac{3\MH^2}{8\MW} I_{020} -\frac{3\MH^2}{2\MW}
I_{121}(\xi)\right)\hH\tr{\hV_\mu \hV^\mu}\nl
+\fac \Bigg(\frac{1}{4}I_{010}
+\MW^2 I_{011}(1)-
I_{111}(1)\Bigg)\tr{\hV_\mu \hV^\mu}\nl
+\fac g^2\left(-\frac{1}{2}I_{112}(\xi) +2
I_{213}(\xi)\right)\tr{\hW_{\mu\nu}\hW^{\mu\nu}}
\nl
+\fac  \bigg(-I_{011}(\xi)+5 I_{112}(\xi)-12
I_{213}(\xi)\bigg)\tr{(\DW^\mu\hV_\mu)^2}\nl
+\fac ig \left(\frac{1}{2}I_{112}(\xi)-4I_{213}(\xi)\right)
 \tr{\hW_{\mu\nu}[\hV^\mu,\hV^\nu]}
\nl
+ \fac \bigg(-4 I_{213}(\xi)+2I_{222}(\xi)\bigg)
\tr{\hV_\mu \hV_\nu}\tr{\hV^\mu \hV^\nu}\nl
+\fac \bigg(\frac{1}{16}I_{020}-\frac{1}{2}I_{121}(\xi)
+4I_{213}(\xi)+
I_{222}(\xi)\bigg)\left(\tr{\hV_\mu \hV^\mu}\right)^2\nl
+\ord{-2},
\label{leff5}
\eeqar
which is manifestly
invariant under gauge transformations of the
background fields \refeq{bgtrafo},
because $\hW^{\mu\nu}$, $\hV^\mu$ and $\DW^\mu \hV_\mu$ transform
covariantly under \refeq{bgtrafo}, i.e.\
\beq
\hW^{\mu\nu} \to S\hW^{\mu\nu}S^\dagger,\qquad
\hV^\mu \to S \hV^\mu S^\dagger,\qquad
\DW^\mu \hV_\mu \to S (\DW^\mu \hV_\mu) S^\dagger.
\label{Vtrafo}
\eeq
In this Lagrangian the gauge for
the background fields can now be fixed arbitraily, e.g.\ in the
$R_\xi$-gauge.

\section{Renormalization}
\label{ren}
\eqnew

In the previous sections we dealt with bare parameters and bare fields
only. Now, we apply the following renormalization transformation to the
parameters
\beqar
g     &\to &          g_0 = g+\de g, \nn\\
\MW^2 &\to & {\MW^2}_{,0} = \MW^2 + \delta \MW^2, \nn\\
\MH^2 &\to & {\MH^2}_{,0} = \MH^2 + \delta \MH^2 , \nn\\
t     &\to &          t_0 = t + \delta t,
\label{parren}
\eeqar
where bare quantities are marked by a subscript ``0'' in the following.
The tadpole term $t = v (\mu^2 - \la v^2/4)$ is introduced via the term
$t\hat H(x)$ in the Lagrangian (\ref{glsmnl}).%
\footnote{Strictly speaking, the relations between the parameters given
in (\ref{phipara}) and (\ref{vmula}) hold for renormalized quantities.
We should have taken a non-vanishing tadpole term $t$ into account for
the unrenormalized parameters. Instead, we omitted $t$ there in order to
avoid confusion, but reintroduce it here.}
Consequently,
the parameter
counterterms are generated by the replacements
\beqar
g     &\to &          g_0 = g+\de g, \nn\\
v     &\to &          v_0 = v+\de v, \nn\\
\mu^2 &\to &      \mu^2_0 = \mu^2+\de\mu^2, \nn\\
\la   &\to &        \la_0 = \la+\de\la,
\label{parren2}
\eeqar
in the Lagrangian (\ref{glsmnl}), where
\beqar
\frac{\de v}{v} &=&
\frac{1}{2} \frac{\de\MW^2}{\MW^2}-\frac{\de g}{g}, \nn\\
\frac{\de\mu^2}{\mu^2} &=&
\frac{\de\MH^2}{\MH^2}+\frac{3g}{2\MW\MH^2}\de t, \nn\\
\frac{\de\la}{\la} &=&
\frac{\de\MH^2}{\MH^2}-\frac{\de\MW^2}{\MW^2}+2\frac{\de g}{g}
+\frac{g}{2\MW\MH^2}\de t.
\label{parctrel}
\eeqar
The masses $\MW$ and $\MH$ are fixed within the on-shell
renormalization scheme, which yields
the usual conditions
(see e.g.\ \citeres{ao80,de93})
\beqar
\de\MW^2 &=& \Re\left\{\Si^{\hW\hW}_{\rm T}(\MW^2)\right\},\nn\\
\de\MH^2 &=& \Re\left\{\Si^{\hH\hH}(\MH^2)\right\},
\label{massct}
\eeqar
where $\Si_{\rm T}^{\hW\hW}$ and $\Si^{\hH\hH}$ represent
the transversal part of the unrenormalized $\hW$ self-energy and the
unrenormalized $\hH$ self-energy, respectively. Concerning the
notation of self-energies, vertex functions etc.\ we follow
\citeres{bfm4,bfm5} throughout. The tadpole counterterm is defined such
that the renormalized tadpole vanishes,
\beq
\de  t= - T^{\hH}.
\eeq
In the following it turns out that the renormalization condition for the
coupling $g$ does not need to be explicitly specified except for the
requirement that it must be defined at a low-energy scale $q^2$, i.e.\
$\vert q^2\vert\ll\MH^2$.
We mention that -- owing to the gauge invariance of the Higgs field and its
vacuum expectation value $v$ -- {\it all} parameter counterterms in the
on-shell
scheme are gauge-independent, i.e.\ independent of $\xi$, which
is in contrast to the situation of a linearly realized Higgs sector
\cite{bfm5}.
Since $\de\MW^2$, $\de t$, $\de g$ are calculated from vertex functions
at low-energy scales, their contributions arising from virtual Higgs
exchange can be directly read from the effective Lagrangian
(\ref{leff5}), yielding
\beqar
\de\MW^2 &=& \fac g^2\left(\frac{1}{4}I_{010}-I_{111}(1)\right)
+\O(\MH^0), \nn\\
\de t &=& -\fac g \frac{3\MH^2}{4\MW}I_{010} + \O(\MH^0), \nn\\
\de g &=& \O(\MH^0).
\label{dmwdtdg}
\eeqar
By definition the physical Higgs mass $\MH$ is fixed at
$q^2=\MH^2\gg\MW^2$ so that the counterterm $\de\MH^2$ has to be
calculated diagrammatically as usual
and cannot be read from \refeq{leff5}. This
is due to the fact that we apply on-shell renormalization. If one
used
a renormalization scheme in that
$\delta\MH^2$ is fixed at $|q^2|\ll\MH^2$, also this renormalization
constant could be calculated by functional methods. However, in
such a renormalization scheme, $\MH^2$ is not the physical Higgs mass.
Thus, in
order to construct the heavy-Higgs limit of the GNLSM, we find it most
consistent to apply on-shell renormalization, where $\MH^2$ is the
physical Higgs mass, although this means that a very small part of our
calculation has to be done diagrammatically.
Since each background Higgs field contributes two inverse
powers of $\MH$ (see \refeq{hm2}),
$\Re\left\{\Si^{\hH\hH}(\MH^2)\right\}$
in (\ref{massct}) has only to be evaluated in
order $\O(\MH^4)$. Hence, we just have to consider the diagrams shown in
\reffi{dmhdiag}.
\begin{figure}
\begin{center}
\begin{picture}(12,2)
\put(-6.5,-15.0){\includegraphics{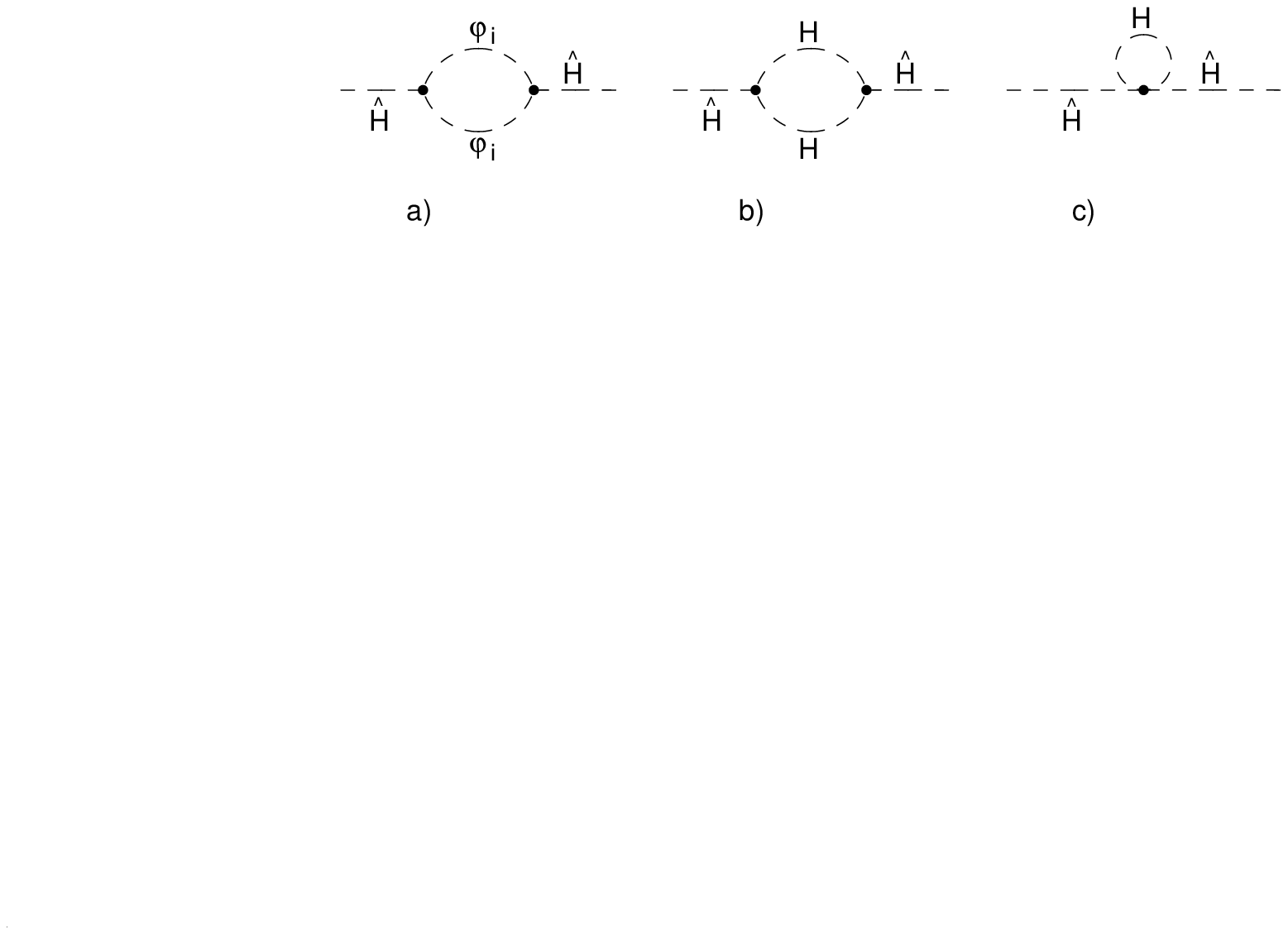}}
\end{picture}
\end{center}
\caption{All diagrams of order $\O(\MH^4)$ in $\de\MH^2$.}
\label{dmhdiag}
\efi
We obtain
\beqar
\de\MH^2 &=& \frac{1}{16\pi^2}g^2\,
\frac{3\MH^2}{8\MW^2} \left[
\MH^2\Re\left\{B_0(\MH^2,0,0)\right\}
+3\MH^2 B_0(\MH^2,\MH,\MH)
+I_{010}
\right]
\nn\\
&&{}+\O(\MH^2), \hspace{2em}
\label{dmh}
\eeqar
where $B_0$ denotes the general scalar two-point function,
\beq
B_0(k^2,M_0,M_1)=
\frac{(2\pi\mu)^{4-D}}{i\pi^2}\int d^D p\;
\frac{1} {[p^2-M_0^2+i\varepsilon] [(p+k)^2-M_1^2+i\varepsilon]}.
\label{b0}
\eeq
The explicit expressions for the $B_0$-terms occurring in (\ref{dmh})
are given in \refapp{ints}.
It should be noted, that in the non-linear parametrization of the
GLSM, the $\hH \vp\vp$-vertex is different from that in the linear one.
Nevertheless, the contribution of the Goldstone loop to
$\delta\MH^2$ in this order is the same in both cases.

{}Finally, we introduce the field renormalization
\beqar
\hW^{\pm} &\to& \hW_{0}^{\pm} = Z_{\hW}^{1/2} \hW^{\pm}
= (1+\frac{1}{2}\de Z_{\hW})\hW^{\pm}, \nn\\
\hvp^{\pm} &\to& \hvp_{0}^{\pm} = Z_{\hvp}^{1/2} \hvp^{\pm}
= (1+\frac{1}{2}\de Z_{\hvp}) \hvp^{\pm}, \nn\\
\hH^{\pm} &\to& \hH_{0} = Z_{\hH}^{1/2} \hH
= (1+\frac{1}{2}\de Z_{\hH}) \hH,
\label{fieren}
\eeqar
for the background fields. In \citere{bfm5} it was demonstrated for the
standard model that field renormalization constants can be chosen such
that in the BFM the renormalized vertex functions obey the same Ward
identities as the unrenormalized ones. Although this will be not of
great importance in view of the heavy-Higgs limit, we note that this
fact also holds in the non-linearly parametrized theory. Inspecting
(\ref{wpara}) and (\ref{U}), one immediately concludes that the
renormalized Lagrangian remains gauge-invariant if
\beqar
\de Z_{\hW}  &=& -2\frac{\de g}{g}, \nn\\
\de Z_{\hvp} &=& 2\frac{\de v}{v},
\label{zwzp}
\eeqar
which is equivalent to the requirement that the renormalized and
unrenormalized vertex functions obey the same Ward identities.
Since in
the non-linear parametrization the Higgs field $\hH$ is an SU(2)
singlet, the $\hH$ field-renormalization constant is not determined
by this condition.
On the other hand, $Z_{\hH}$ will drop out
anyhow if $\hH$ is removed so that we can simply choose
\beq
Z_{\hH}=1,
\qquad \de Z_{\hH}=0.
\label{zh}
\eeq
In this context, we mention that the off-shell self-energy
$\Si^{\hH\hH}(k^2)$ of $\hH$ will even be UV-divergent
for any value of $\de Z_{\hH}$ owing to the presence of UV-divergent
terms proportional to $k^4$. Of course, the occurrence of such a term is
an artefact of the non-linear parametrization of the Higgs sector since
the complete theory is nevertheless renormalizable.

Applying finally the complete renormalization transformation
(\ref{parren}), (\ref{fieren}) to the
Lagrangian (\ref{glsmnl}), we obtain the $\hH$-dependent part
$\lcth$ of the counterterm Lagrangian,
\beq
\lcth =
\de t\hH - \frac{1}{2}\de\MH^2\hH^2
- \frac{1}{2}\frac{\de\MW ^2}{g\MW}\hH\tr{\hV^\mu\hV_\mu} + \O(\zeta^{-2}).
\eeq
In slight abuse of terminology we call the sum of the effective
Lagrangian $\leff$ and $\lcth$ the {\it renormalized effective Lagrangian}
$\lreff$. With the explicit expressions of
\refeq{leff5},
(\ref{dmwdtdg}), (\ref{dmh}),
(\ref{zwzp}), and (\ref{zh}) we get the result
\beqar
\lreff&=&
\spc\fac \frac{3g^2\MH^4}{16\MW^2} \bigg(
3I_{020}-3B_0(\MH^2,\MH,\MH)-\Re\left\{B_0(\MH^2,0,0)\right\}
\bigg)\hH^2
\nl
+\fac \frac{g}{8\MW}
\bigg(-I_{010}+4I_{111}(1)+3\MH^2I_{020}
-12\MH^2I_{121}(\xi)\bigg)\hH\tr{\hV_\mu \hV^\mu}
\nl
+\fac \bigg(\frac{1}{4}I_{010}
+\MW^2 I_{011}(1)-
I_{111}(1)\bigg)\tr{\hV_\mu \hV^\mu}\nl
+\fac g^2\left(-\frac{1}{2}I_{112}(\xi) +2
I_{213}(\xi)\right)\tr{\hW_{\mu\nu}\hW^{\mu\nu}}
\nl
+\fac  \bigg(-I_{011}(\xi)+5 I_{112}(\xi)-12
I_{213}(\xi)\bigg)\tr{(\DW^\mu\hV_\mu)^2}\nl
+\fac ig  \left(\frac{1}{2}I_{112}(\xi)-4I_{213}(\xi)\right)
 \tr{\hW_{\mu\nu}[\hV^\mu,\hV^\nu]}
\nl
+ \fac \bigg(-4 I_{213}(\xi)+2I_{222}(\xi)\bigg)
\tr{\hV_\mu \hV_\nu}\tr{\hV^\mu \hV^\nu}\nl
+\fac \bigg(\frac{1}{16}I_{020}-\frac{1}{2}I_{121}(\xi)
+4I_{213}(\xi)+
I_{222}(\xi)\bigg)\left(\tr{\hV_\mu \hV^\mu}\right)^2\nl
+\ord{-2},
\label{leff6}
\eeqar
where the (tadpole) terms linear in $\hH$ drop out as expected.

\section{Elimination of the background Higgs field}
\label{bghelim}
\eqnew
After having integrated out the quantum Higgs field $H$, the
effective Lagrangian \refeq{leff6} still contains the background
Higgs field $\hH$. Integrating out the quantum Higgs corresponds
in the diagrammatical formalism to the
calculation of the effects of the heavy Higgs boson in loops.
The elimination of the background Higgs field yields the
effects of the Higgs field outside loops.

The background fields occur as tree lines in the diagrammatical
calculation of (reducible) Green functions and S-matrix elements.
For $\MH\to\infty$ no external $\hH$ fields occur, and (internal)
$\hH$ propagators can be expanded in powers of $1/\MH^2$.
Diagrammatically this means that the $\hH$ propagator shrinks to a point
rendering such (sub-)graphs irreducible which contain $\hH$ lines only.
Inspecting the $\hH$-terms in the Lagrangian of the GLSM, one easily
finds that this expansion leads to the replacement
\beq
\hH\to -\frac{\MW}{g\MH^2}\tr{\hV_\mu\hV^\mu} +\OH{-4}.
\label{heomex}
\eeq
This substitution can also be motivated as follows.
The tree-like parts of Feynman graphs correspond to the lowest
order in the perturbative expansion of amplitudes which is known to
agree with the lowest-order result of the classical equations of motion
(EOM). The EOM for the $\hH$ field is given by
\beq
(\partial^2 +\MH^2)\hH =-\frac{\MW}{g} \tr{\hV_\mu\hV^\mu}
-\frac{1}{2} \hH\tr{\hV_\mu\hV^\mu}-\frac{3g\MH^2}{4\MW}\hH^2
-\frac{g^2\MH^2}{8\MW^2}\hH^3,
\label{heom}
\eeq
which can be solved for $\hH$ by recursion. The leading contribution
exactly corresponds to \refeq{heomex}.

{}First, we insert \refeq{heomex} into the tree-level Lagrangian of the
GLSM, i.e.\ that part of \refeq{glsmnl} that contains only background
fields:
\beqar
\lag^{\rm tree}&=& -\frac{1}{2} \tr{\hW_{\mu\nu}\hW^{\mu\nu}} + \frac{1}{2}
(\partial_\mu \hH)(\partial^\mu \hH) - \frac{1}{4}
\left(\frac{2\MW}{g}+\hH\right)^2\tr{\hV_\mu \hV^\mu } \nn\\
&&{}-\frac{1}{2}\MH^2\hH^2-\frac{g^2\MH^2}{4\MW}\hH^3
-\frac{g^2\MH^2}{32\MW^2}\hH^4.
\label{ltree}\eeqar
One immediately finds that in $\OH{0}$ \refeq{heomex} results in simply
dropping $\hH$ in \refeq{ltree}. This is the well-known result that the
limit for $\MH\to\infty$ of the gauged linear $\sigma$-model
at tree level is the gauged non-linear $\sigma$-model \cite{apbe}:
\beq
\lag^{\rm tree}|_{\MH\to\infty}=\lag^{\rm tree}|_{\hH=0}+\OH{-2}
=\lag^{\rm tree}_{\rm GNLSM}+\OH{-2},
\eeq
with
\beq
\lag^{\rm tree}_{\rm GNLSM}=
-\frac{1}{2} \tr{\hW_{\mu\nu}\hW^{\mu\nu}}-
\frac{\MW^2}{g^2}\tr{\hV_\mu \hV^\mu }.
\label{lgnlsmtree}
\eeq

Eq.~\refeq{pi2} implies that the one-loop Lagrangian of the GLSM for
$\MH\to\infty$ consists of two parts, namely the effective Lagrangian
$\lreff$ of \refeq{leff6}
and the remainder $\lag^{\rm 1-loop}|_{H=0}$ in the
path integral \refeq{pi2}. The effective Lagrangian
was generated by integrating out the quantum Higgs, i.e.\ it
parametrizes the effects of loops containing the heavy Higgs bosons.
The remainder Lagrangian $\lag^{\rm 1-loop}|_{H=0}$
still contains the light quantum
fields, i.e.\ it has to be used in order to calculate the
contributions from loops without the heavy Higgs boson.
As in the case of the tree-level
Lagrangian, in $\OH{0}$ the application
of substitution \refeq{heomex}
simply results in dropping $\hH$ in $\lag^{\rm 1-loop}|_{H=0}$,
which yields the one-loop Lagrangian of GNLSM.
The effective Lagrangian \refeq{leff6}, however, parametrizes  the
differences of the GLSM for $\MH\to\infty$ and the GNLSM.
Applying \refeq{heomex} to
\refeq{leff6} yields
non-vanishing
contributions at $\OH{0}$. One finds
\beqar
\lreff&=&\spc\fac \bigg(\frac{1}{4}I_{010}
+\MW^2 I_{011}(1)-
I_{111}(1)\bigg)\tr{\hV_\mu \hV^\mu}\nl
+\fac \bigg(-\frac{1}{2}I_{112}(\xi) +2
I_{213}(\xi)\bigg)g^2\tr{\hW_{\mu\nu}\hW^{\mu\nu}}
\nl
+\fac  \bigg(-I_{011}(\xi)+5 I_{112}(\xi)-12
I_{213}(\xi)\bigg)\tr{(\DW^\mu\hV_\mu)^2}\nl
+\fac  \bigg(\frac{1}{2}I_{112}(\xi)-4I_{213}(\xi)\bigg)ig
 \tr{\hW_{\mu\nu}[\hV^\mu,\hV^\nu]}
\nl
+ \fac \bigg(-4 I_{213}(\xi)+2I_{222}(\xi)\bigg)
\tr{\hV_\mu \hV_\nu}\tr{\hV^\mu \hV^\nu}\nl
+\fac\bigg(
\frac{1}{8\MH^2}I_{010}-\frac{1}{2\MH^2}I_{111}(1)
+\frac{1}{4}I_{020}+I_{121}(\xi)
+4I_{213}(\xi)+
I_{222}(\xi)
\nl\qquad\qquad\qquad
-\frac{9}{16}B_0(\MH^2,\MH,\MH)
-\frac{3}{16}\Re\left\{B_0(\MH^2,0,0)\right\}
\bigg)\left(\tr{\hV_\mu \hV^\mu}\right)^2\nl
+\O(\MH^{-2}).
\label{leff7}
\eeqar

Inserting the explicit expressions \refeq{integrals} and \refeq{Bs}
for the integrals occurring in \refeq{leff7}
we finally find for $\lreff$:
\beqar
\lreff&=&\spc\fac \Bigg[-\frac{1}{8}\MH^2 +\frac{3}{4}\MW^2\left(
\tde+\frac{5}{6}\right)
\Bigg]\tr{\hV_\mu \hV^\mu}\nl
-\fac \frac{1}{24}\left(\tde+\frac{5}{6}
\right)g^2\tr{\hW_{\mu\nu}\hW^{\mu\nu}}
\nl
{}-\fac  \frac{1}{4}\left(\tde+\frac{1}{6}
\right)\tr{(\DW^\mu\hV_\mu)^2}\nl
-\fac \frac{1}{24} \left(\tde+\frac{17}{6}\right)
 ig \tr{\hW_{\mu\nu}[\hV^\mu,\hV^\nu]}
\nl
- \fac \frac{1}{12}\left(\tde+\frac{17}{6}\right)
\tr{\hV_\mu \hV_\nu}\tr{\hV^\mu \hV^\nu}\nl
-\fac \frac{1}{24}\left(\tde +\frac{79}{3}-\frac{27\pi}{2\sqrt{3}}
\right)
\left(\tr{\hV_\mu \hV^\mu}\right)^2\nl
+\O(\MH^{-2}),
\label{lefffinal}
\eeqar
with $\tde$ being given in \refeq{tde}.

As a final result for the one-loop Lagrangian of the GLSM
for $\MH\to\infty$ we find:
\beqar
\lag^{\rm 1-loop,ren}|_{\MH\to\infty} &=&
\lag^{\rm 1-loop}|_{H=\hH=0}+
\lag^{\rm ct}|_{\hH=0}+
\lreff+\OH{-2} \nn\\[.3em]
&=& \lag^{\rm 1-loop}_{\rm GNLSM}+
\lag^{\rm ct}|_{\hH=0}+\lreff+\OH{-2}.
\label{lfinal}
\eeqar

In (\ref{lfinal}) the counterterm Lagrangian $\lag^{\rm ct}|_{\hH=0}$
simply follows from the tree-level GNLSM
Lagrangian (\ref{lgnlsmtree}) upon applying the renormalization
transformations for $v$, $\MW$, $g$, $\hvp$, and $\hW$
(see (\ref{parren}), (\ref{parren2}), (\ref{parctrel}), and
(\ref{fieren})), where $\de\MW^2$ is fixed by the on-shell condition
(\ref{massct}), and $\de Z_{\hvp}$ by (\ref{zwzp}). We did not
make use of a specification for $\de g$ and $\de Z_{\hW}$ so that these
counterterms can be chosen independently. Note that the contributions of
the effective Lagrangian $\lreff$ (\ref{lfinal}) have to be included in
the determination of the counterterms.

The effective Lagrangian $\lreff$ of (\ref{lefffinal}) quantifies the
exact difference between the heavy-Higgs GLSM (in non-linear
representation of the Higgs sector) and the GNLSM at one loop.
More precisely, applying (\ref{lefffinal}) one obtains the difference
for each vertex function, where the tadpole and Higgs-mass
renormalization has already been carried out in the GLSM, but the
remaining renormalization is still to be done. Inspecting
(\ref{lefffinal}), one finds that the first two terms have the same
structure as terms in the tree-level Lagrangian of the GNLSM
(\ref{lgnlsmtree}).
Thus, the contributions of the $\tr{\hV_\mu \hV^\mu}$ and
$\tr{\hW_{\mu\nu}\hW^{\mu\nu}}$-terms in \refeq{lefffinal} can be
absorbed into the corresponding counterterms. This means that S-matrix
elements are not influenced by these terms.

{}Furthermore,
it should be noted that the $\tr{(\DW^\mu\hV_\mu)^2}$-term in
\refeq{lefffinal} is redundant
in view of the calculation of (reducible) Green functions and S-matrix
elements.
Actually, we may not only use the EOM for $\hH$ but also those for
$\hW^\mu$ in order to simplify the effective interaction
term $\lreff$, although in the latter case this does not correspond
to a $1/\MH$-expansion. The reason for this is that
the application of the lowest-order EOM in
$\lreff$ simply corresponds to a field transformation of the
background fields, which leaves S-matrix elements invariant,
and to an expansion in the coupling constant of the effective interaction
term (i.e.\ in $g^2/(16\pi^2)$ in our case) \cite{eom}.
However, the EOM for the vector fields in the GNLSM
\beq
\DW^\mu \hW_{\mu\nu} = -\frac{i}{g}\MW^2\hV_\nu
\eeq
implies
\beq
\DW^\mu \hV_\mu=0.
\label{dv0}\eeq
Hence, the contributions of the $\tr{(\DW^\mu\hV_\mu)^2}$-term drop out
in complete (reducible) Green functions and S-matrix elements, even
though this term yields non-vanishing contributions to single vertex
functions. Consequently, the effective Lagrangian
\beqar
\lreff(\mbox{S-matrix}) &=&
-\fac \frac{1}{24} \left(\tde+\frac{17}{6}\right)
 ig \tr{\hW_{\mu\nu}[\hV^\mu,\hV^\nu]}
\nl
- \fac \frac{1}{12}\left(\tde+\frac{17}{6}\right)
\tr{\hV_\mu \hV_\nu}\tr{\hV^\mu \hV^\nu}\nl
-\fac \frac{1}{24}\left(\tde +\frac{79}{3}-\frac{27\pi}{2\sqrt{3}}
\right)
\left(\tr{\hV_\mu \hV^\mu}\right)^2\nl
+\O(\MH^{-2})
\label{leffsmatrix}
\eeqar
summarizes the complete differences between the GLSM and the GNLSM
contributing to
the S-matrix.

\section{Some examples of vertex functions}
\label{examples}
\eqnew
In order to illustrate our results, we consider some special vertex
functions and calculate their difference between the GLSM with a heavy
Higgs boson and the corresponding GNLSM. For these examples we compare
the results derived from the effective Lagrangian (\ref{leff7})
with the
ones obtained by evaluating directly the Feynman diagrams.
Note that if the non-linear parametrization, \refeq{phinl} with
\refeq{U}, is used for the GLSM, the difference between the GLSM and
the GNLSM in diagrammatical calculations comes only from diagrams
with internal Higgs lines. The situation
will be different if the linear
para\-metrization \refeq{philin} is used, as in \citere{hemo}, where
also some diagrams without Higgs lines differ in both models.

Instead of
using the fields $W_i$ and $\vp_i$ as introduced in \refse{bfmstf}, we
find it convenient to introduce the fields
\beq
\begin{array}[b]{rclcrcl}
W_\mu^\pm &=& \disp\frac{1}{\sqrt{2}}\left(W_\mu^1\mp i W_\mu^2\right),
&\qquad & W_\mu^0 &=& W_{\mu}^3, \\[1em]
\vp^\pm   &=& \disp\frac{1}{\sqrt{2}}\left(\vp_2\pm i\vp_1\right),
&&        \chi    &=& -\vp_3.
\earr
\label{phyfie}
\eeq
We mention that the definitions of $W_\mu^\pm$, $\vp^\pm$, and $\chi$
follow the ones for the SM fields of
\citeres{bfm4,bfm5,de93}, where the linear parametrization (\ref{philin})
is used. The SM $Z$-field introduced there reduces to our $W^0$-field
upon replacing the couplings $g_2\to g$, $g_1\to 0$.

We start by considering the $\hW^0$ self-energy. In \reffi{w0se} we show
the corresponding Feynman graphs which contain the Higgs field in the
GLSM, but are absent in the GNLSM.
As mentioned above,
the graphs of
\reffi{w0se} form exactly the difference between the GLSM (with
non-linear Higgs realization) and the GNLSM. We mention that we have not
explicitly written down the tadpole graph and its counterterm, since
these terms obviously cancel. Calculation of each graph yields
\beqar
\Ga^{\hW^0\hW^0}_{\mu\nu}(k_0)\Big\vert_{\ref{w0se}\rm a)} &=&
\disp\mu^{4-D}\int\frac{d^D p}{(2\pi)^D}\;
\frac{(-g^2)\MW^2}{p^2-\MH^2} \nn\\
&& \hspace{3em} \times
\left\{ \frac{g_{\mu\nu}}{(p+k_0)^2-\MW^2}
       -\frac{(1-\xi)(p+k_0)_\mu(p+k_0)_\nu}
         {\left[(p+k_0)^2-\MW^2\right]\left[(p+k_0)^2-\xi\MW^2\right]}
\right\} \nn\\
&=& -\frac{ig^2}{16\pi^2}g_{\mu\nu}
\left\{ \MW^2I_{011}(1)+I_{111}(\xi)-I_{111}(1) \right\} +\OH{-2},
\nn\\[.5em]
\Ga^{\hW^0\hW^0}_{\mu\nu}(k_0)\Big\vert_{\ref{w0se}\rm b)} &=&
\disp\mu^{4-D}\int\frac{d^D p}{(2\pi)^D}\;
\frac{g^2(p+k_0)_\mu(p+k_0)_\nu}
     {\left[p^2-\MH^2\right]\left[(p+k_0)^2-\xi\MW^2\right]}
\nn\\
&=& \frac{ig^2}{16\pi^2} \Big\{
g_{\mu\nu}\left[I_{111}(\xi)-k_0^2I_{112}(\xi)+4k_0^2I_{213}(\xi)\right]
\nn\\
&& \phantom{\frac{ig^2}{16\pi^2} \Big\{}
+k_{0,\mu}k_{0,\nu}\left[I_{011}(\xi)-4I_{112}(\xi)+8I_{213}(\xi)\right]
\Big\} +\OH{-2},
\nn\\[.5em]
\Ga^{\hW^0\hW^0}_{\mu\nu}(k_0)\Big\vert_{\ref{w0se}\rm c)} &=&
\disp\mu^{4-D}\int\frac{d^D p}{(2\pi)^D}\;
\frac{(-g^2)g_{\mu\nu}}{4\left(p^2-\MH^2\right)}
\nn\\
&=& -\frac{ig^2}{16\pi^2} \frac{1}{4}g_{\mu\nu}I_{010},
\eeqar
where we have already performed the
$1/\MH$-expansions
by expanding the Feynman integrals around the vacuum integrals defined
in \refeq{intnot}.
Adding up terms
\ref{w0se}a)--\ref{w0se}c), one obtains
\beqar
\de\Ga^{\hW^0\hW^0}_{\mu\nu}(k_0) &=&
\frac{ig^2}{16\pi^2} \Biggl\{
g_{\mu\nu}\left[I_{111}(1)-\MW^2I_{011}(1)-\frac{1}{4}I_{010}
                -k_0^2I_{112}(\xi)+4k_0^2I_{213}(\xi)\right]
\nn\\
&& \phantom{\frac{ig^2}{16\pi^2} \Biggl\{}
+k_{0,\mu}k_{0,\nu}\left[I_{011}(\xi)-4I_{112}(\xi)+8I_{213}(\xi)\right]
\Biggr\} +\OH{-2}.
\label{w0w0res}
\eeqar
This result has to be compared with the contribution of the
effective Lagrangian (\ref{leff7}) to the $\hW^0$ self-energy.
To this end, we give the $\hW^0\hW^0$-parts
of the relevant traces
occurring in (\ref{leff7}) so that one can easily read off
the Feynman rules,
\beqar
\tr{\hW_{\mu\nu}\hW^{\mu\nu}}\Big\vert_{\hW^0\hW^0} &=&
-(\partial\hW^0)^2-\hW^0_\mu\partial^2\hW^{0,\mu}, \nn\\
\tr{(\DW^\mu\hV_\mu)^2}\Big\vert_{\hW^0\hW^0} &=&
-\frac{g^2}{2}(\partial\hW^0)^2, \nn\\
\tr{\hV_\mu \hV^\mu}\Big\vert_{\hW^0\hW^0} &=&
-\frac{g^2}{2}(\hW^0)^2.
\label{w0w0leff}
\eeqar
Using (\ref{leff7}) with (\ref{w0w0leff}) one directly obtains
(\ref{w0w0res}).

\begin{figure}
\begin{center}
\begin{picture}(12,2)
\put(-2.5,-15.0){\includegraphics{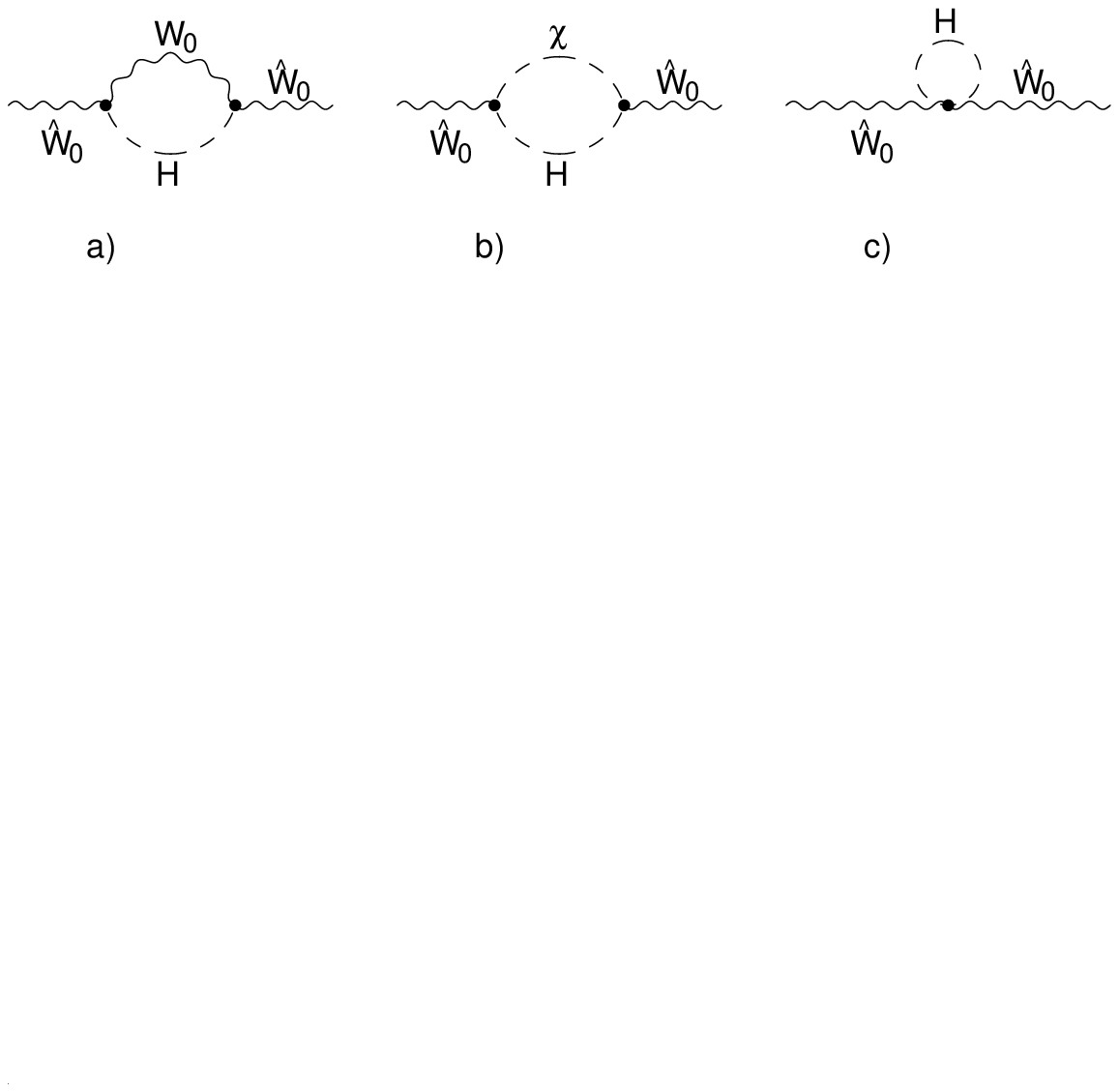}}
\end{picture}
\end{center}
\caption{Higgs diagrams for the $\hW^0$ self-energy.}
\label{w0se}
\efi
\begin{figure}
\begin{center}
\begin{picture}(12,2)
\put(-2.5,-15.0){\includegraphics{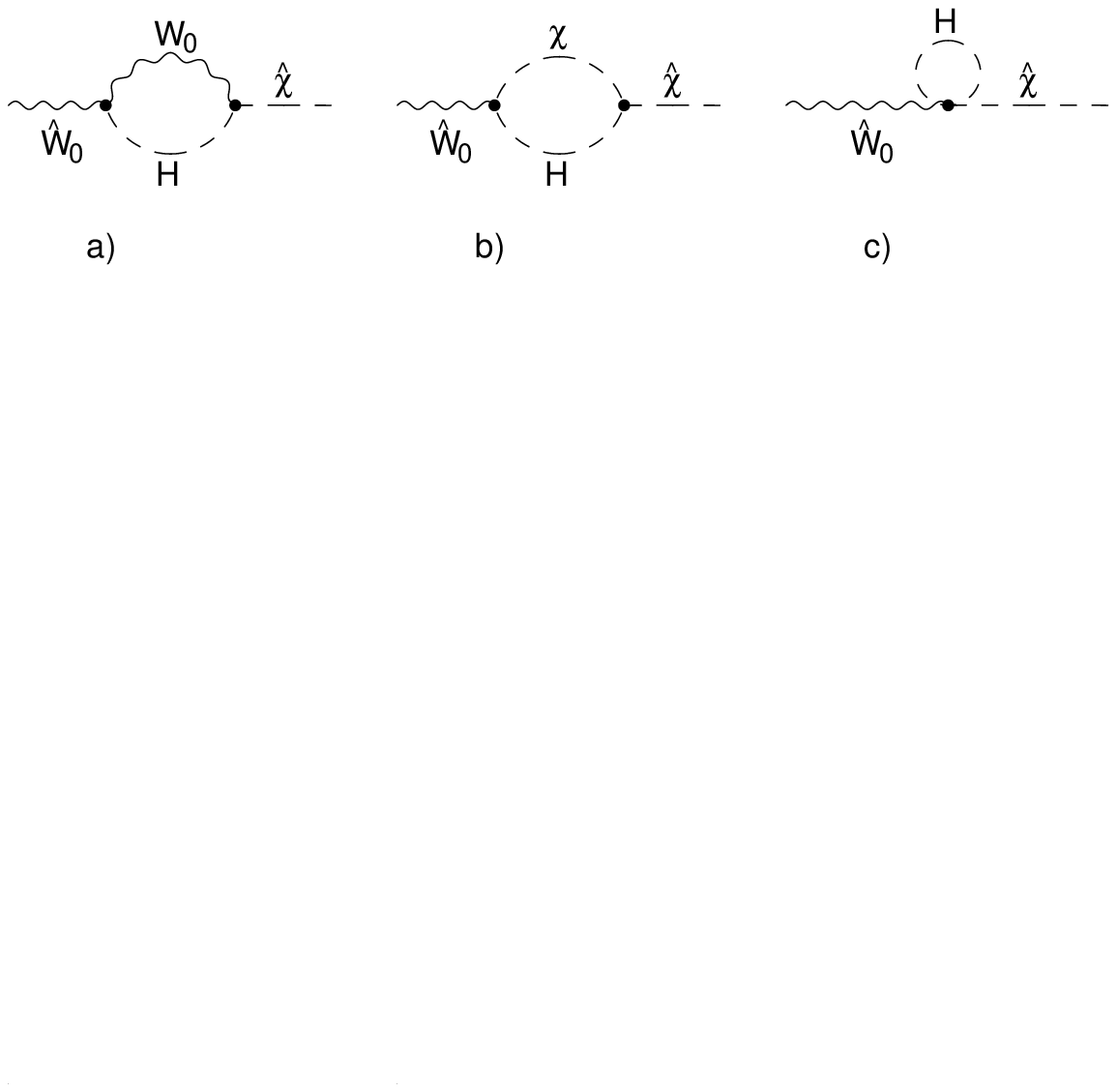}}
\end{picture}
\end{center}
\caption{Higgs diagrams for the $\hW^0\hc$-mixing self-energy.}
\label{w0cse}
\efi
Next, we turn to the $\hW^0\hc$-mixing self-energy. \reffi{w0cse}
shows the diagrams for the difference between the GLSM and the GNLSM in
analogy to the previous example. The tadpoles (which cancel) are again
not explicitly drawn. The individual graphs are calculated to
\beqar
\Ga^{\hW^0\hc}_{\mu}(k_0)\Big\vert_{\ref{w0cse}\rm a)} &=&
\disp\mu^{4-D}\int\frac{d^D p}{(2\pi)^D}\;
\frac{(-ig^2)\MW k_{0,\al}}{p^2-\MH^2} \nn\\
&& \hspace{3em} \times
\left\{ \frac{g_{\mu\al}}{(p+k_0)^2-\MW^2}
       -\frac{(1-\xi)(p+k_0)_\mu(p+k_0)_\al}
         {\left[(p+k_0)^2-\MW^2\right]\left[(p+k_0)^2-\xi\MW^2\right]}
\right\} \nn\\
&=& \frac{g^2}{16\pi^2} \frac{k_{0,\mu}}{\MW}
\left\{ \MW^2I_{011}(1)+I_{111}(\xi)-I_{111}(1) \right\} +\OH{-2},
\nn\\[.5em]
\Ga^{\hW^0\hc}_{\mu}(k_0)\Big\vert_{\ref{w0cse}\rm b)} &=&
\disp\mu^{4-D}\int\frac{d^D p}{(2\pi)^D}\;
\frac{ig^2}{\MW}
\frac{(p+k_0)_\mu(pk_0+k_0^2)}
     {\left[p^2-\MH^2\right]\left[(p+k_0)^2-\xi\MW^2\right]}
\nn\\
&=& -\frac{g^2}{16\pi^2} \frac{k_{0,\mu}}{\MW}
\left\{I_{111}(\xi)+k_0^2I_{011}(1)
       -5k_0^2I_{112}(\xi)+12k_0^2I_{213}(\xi)\right\}
+\OH{-2},
\nn\\[.5em]
\Ga^{\hW^0\hc}_{\mu}(k_0)\Big\vert_{\ref{w0cse}\rm c)} &=&
\disp\mu^{4-D}\int\frac{d^D p}{(2\pi)^D}\;
\frac{(-ig^2)k_{0,\mu}}{4\MW(p^2-\MH^2)}
\nn\\
&=& \frac{g^2}{16\pi^2} \frac{k_{0,\mu}}{4\MW} I_{010}.
\eeqar
The resulting contribution
\beqar
\de\Ga^{\hW^0\hc}_{\mu}(k_0) &=&
\frac{g^2}{16\pi^2} \frac{k_{0,\mu}}{\MW} \Biggl\{
\MW^2I_{011}(1)-I_{111}(1)+\frac{1}{4}I_{010}
\nn\\
&& \phantom{\frac{g^2}{16\pi^2} \frac{k_{0,\mu}}{\MW} \Biggl\{}
-k_0^2I_{011}(\xi)+5k_0^2I_{112}(\xi)-12k_0^2I_{213}(\xi)
\Biggr\} +\OH{-2}
\label{w0cres}
\eeqar
again agrees with the result from the effective Lagrangian
(\ref{leff7}). The necessary Feynman rules for the effective
$\hW^0\hc$-mixing follow by inserting the corresponding terms
\beqar
\tr{\hV_\mu \hV^\mu}\Big\vert_{\hW^0\hc} &=&
-\frac{g^2}{\MW}(\partial^\mu\hc)\hW^0_\mu, \nn\\
\tr{(\DW^\mu\hV_\mu)^2}\Big\vert_{\hW^0\hc} &=&
-\frac{g^2}{\MW}(\partial^2\hc)(\partial\hW^0)
\label{w0cleff}
\eeqar
into (\ref{leff7}), and one immediately obtains (\ref{w0cres}).
We remark that the differences $\de\Ga^{\hW^0\hW^0}_{\mu\nu}$ and
$\de\Ga^{\hW^0\hc}_{\mu}$ obey the Ward identity
\beq
0 = k^\mu_0 \Ga^{\hW^0\hW^0}_{\mu\nu}(k_0)+i\MW\Ga^{\hW^0\hc}_{\nu}(k_0),
\label{w0w0wi}
\eeq
which is fulfilled both in the GLSM with non-linearly realized scalar
sector and the GNSLM.

{}Finally, we inspect the less trivial example of the four-point function
$\Ga^{\hW^+\hW^-\hW^0\hW^0}_{\mu\nu\rho\si}$. The heavy-Higgs
contributions to $\Ga^{\hW^+\hW^-\hW^0\hW^0}_{\mu\nu\rho\si}$ in the
GLSM can be classified into three topologically different types:
irreducible diagrams (\reffi{w4irr}), and reducible diagrams with either
one (\reffi{w4re1}) or two (\reffi{w4re2}) $\hH$ fields on tree lines.
Again all tadpole terms cancel and are omitted from the beginning.%
\footnote{Note that Feynman diagrams with tree lines of background
fields other than $\hH$ give no contributions to (irreducible!) vertex
functions, since only the $\hH$ propagator shrinks to a point for
$\MH\to\infty$.}
In \reffis{w4irr}-\ref{w4re2} we only show the diagrams which are at
least of order $\O(\MH^0)$; the (numerous) graphs of order
$\O(\MH^{-2})$ are omitted.

The irreducible graphs are calculated to
\beqar
\de\Ga^{\hW^+\hW^-\hW^0\hW^0}_{\mu\nu\rho\si}
(k_+,k_-,k_{0,1},k_{0,2})\Big\vert_{\rm irr} &=&
-\frac{ig^4}{16\pi^2}g_{\mu\nu}g_{\rho\si}
\left[I_{121}(\xi)-\frac{1}{8}I_{020}-2I_{222}(\xi)\right] \nn\\
&& +\frac{ig^4}{16\pi^2}(g_{\mu\rho}g_{\nu\si}+g_{\mu\si}g_{\nu\rho})
\,2I_{222}(\xi).
\label{w4irrexp}
\eeqar
Instead of writing down the explicit expression for each diagram, we add
some remarks on the single contributions. Inspecting the momentum
integrals, one finds that the external momenta $k_\pm, k_{0,1/2}$ do not
give contributions of $\OH{0}$ and can be set to zero for all diagrams
of \reffi{w4irr}. Hence, all diagrams can be expressed in terms of
$I_{klm}(\xi)$-functions defined in (\ref{intnot}). The
$I_{213}(\xi)$-terms, which originate from graphs \ref{w4irr}a)-h),
exactly cancel each other.
\begin{figure}
\begin{center}
\begin{picture}(16,11)
\put(-2.5,-6.3){\includegraphics{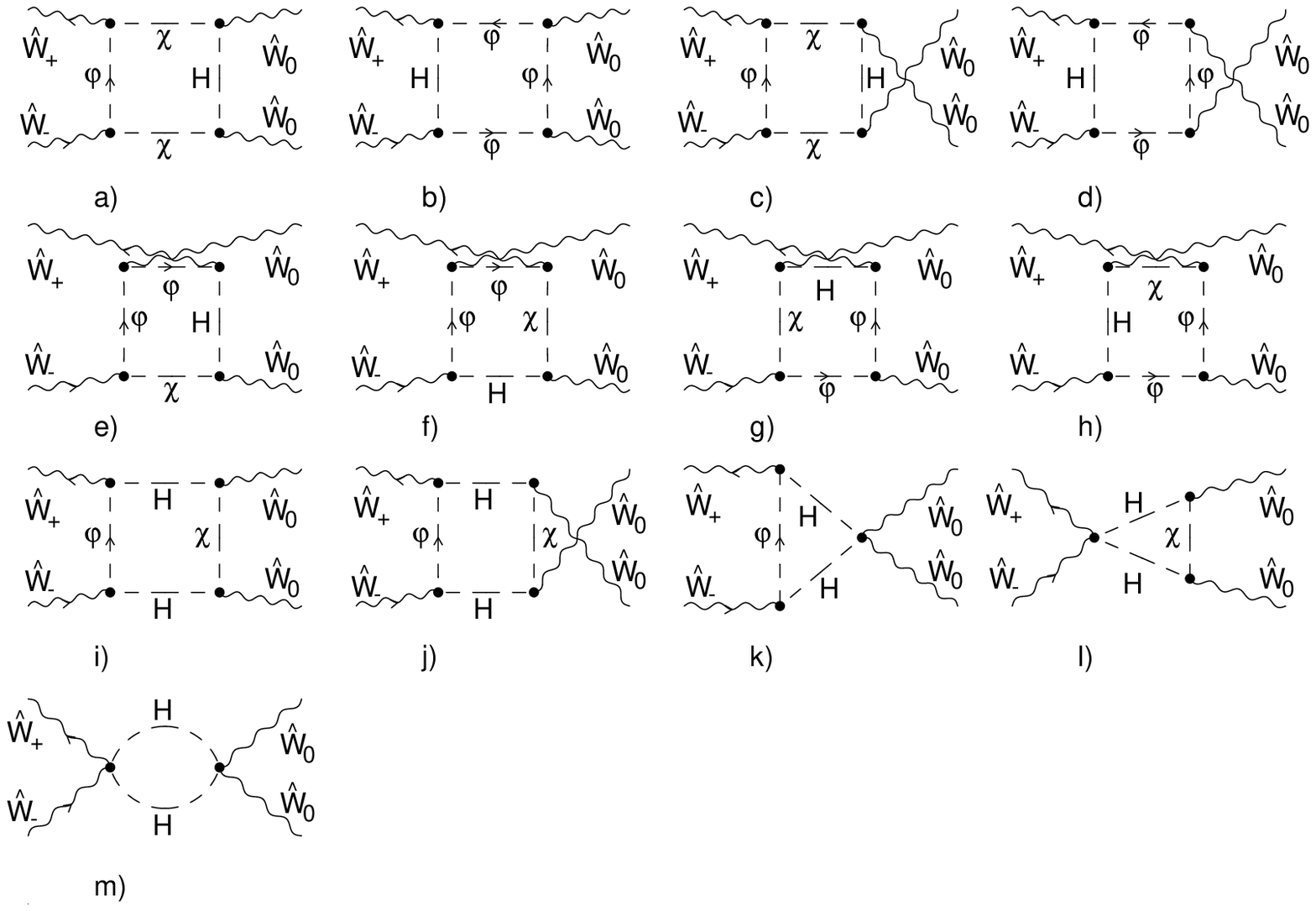}}
\end{picture}
\end{center}
\caption{Irreducible Higgs diagrams for the $\hW^+\hW^-\hW^0\hW^0$
four-point-function.}
\label{w4irr}
\efi

The sum of all diagrams involving exactly one background Higgs
propagator (\reffi{w4re1}) is given by
\beqar
\de\Ga^{\hW^+\hW^-\hW^0\hW^0}_{\mu\nu\rho\si}
(k_+,k_-,k_{0,1},k_{0,2})\Big\vert_{\hH}
&& \nn\\
&& \hspace{-11em}
= \frac{ig^4}{16\pi^2}g_{\mu\nu}g_{\rho\si}
\left[3I_{121}(\xi)-\frac{3}{4}I_{020}
+\frac{1}{\MH^2}\left(\frac{1}{4}I_{010}-I_{111}(\xi)\right)\right].
\hspace{2em}
\label{w4re1exp}
\eeqar
The W-mass counterterm $\de\MW^2$ originates from
the renormalization of the $\hH\hW^+\hW^-$ and $\hH\hW^0\hW^0$
couplings, which is indicated by
the graphs
\ref{w4re1}k),l), and is explicitly given in (\ref{dmwdtdg}). Again, the
external momenta do not contribute in $\OH{0}$. Note also that the
diagrams \ref{w4re1}a),e),f) and \ref{w4re1}c),g),h), which yield
contributions of $I_{111}(\xi)$, cancel each other, respectively.
\begin{figure}
\begin{center}
\begin{picture}(16,8.5)
\put(-2.5,-8.8){\includegraphics{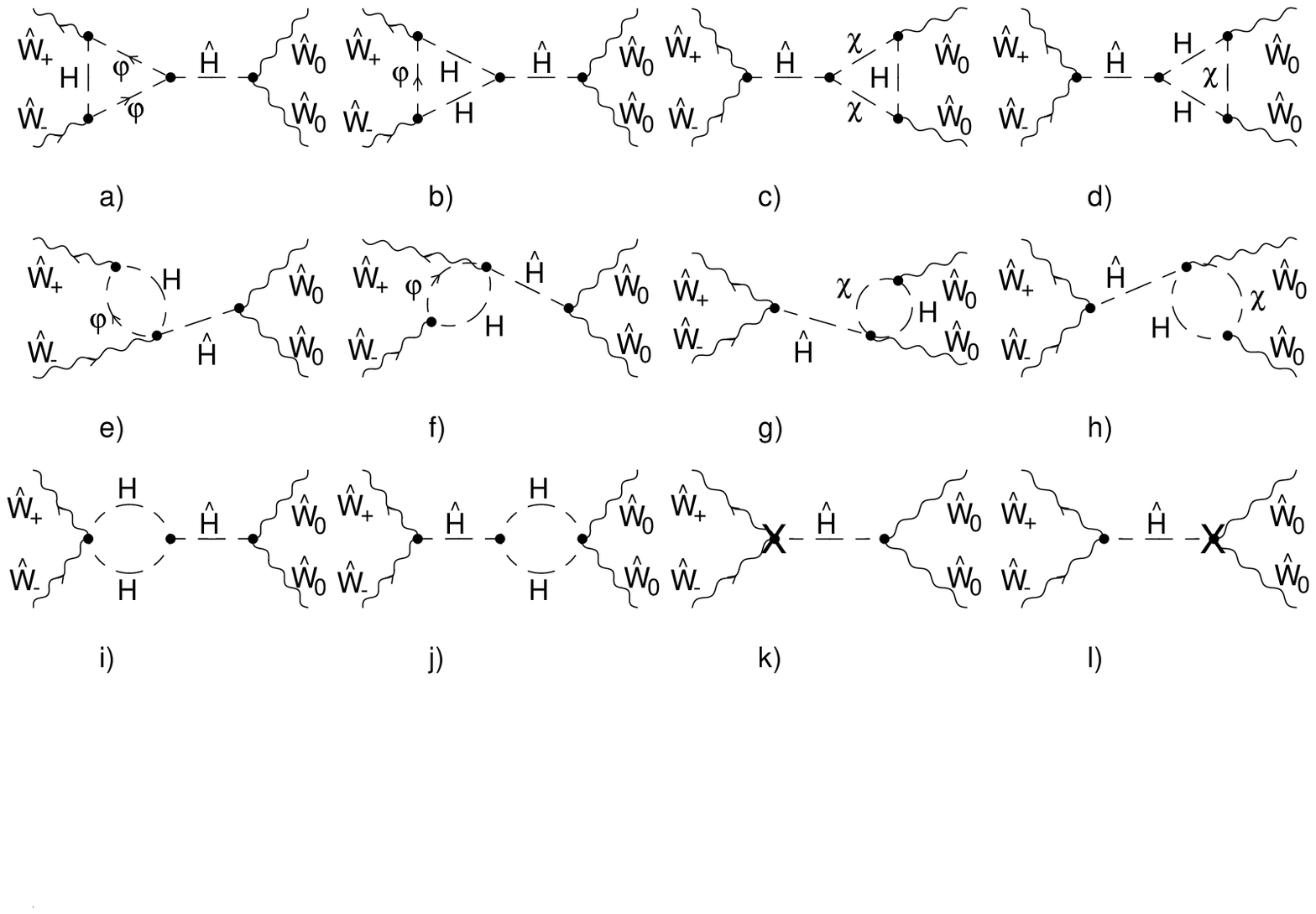}}
\end{picture}
\end{center}
\caption{Reducible Higgs diagrams for the $\hW^+\hW^-\hW^0\hW^0$
four-point-function with {\it one} background Higgs field.}
\label{w4re1}
\efi

The diagrams with two background Higgs propagators (\reffi{w4re2})
represent the contribution of the renormalized Higgs
self-energy $\Sigma^{\hH\hH,{\rm ren}}(q^2)$
at $q^2=(k_++k_-)^2$.
In $\OH{0}$ one only needs $\Sigma^{\hH\hH,{\rm ren}}(0)$ and obtains
\beqar
\de\Ga^{\hW^+\hW^-\hW^0\hW^0}_{\mu\nu\rho\si}
(k_+,k_-,k_{0,1},k_{0,2})\Big\vert_{\hH\hH}
&& \nn\\
&& \hspace{-11em}
= \frac{ig^4}{16\pi^2}g_{\mu\nu}g_{\rho\si}
\left[\frac{9}{8}I_{020}-\frac{9}{8}B_0(\MH^2,\MH,\MH)
-\frac{3}{8}\Re\left\{B_0(\MH^2,0,0)\right\}\right],
\hspace{2em}
\label{w4re2exp}
\eeqar
where the Higgs-mass counterterm can be read from (\ref{dmwdtdg}). Of
course, diagram \ref{w4re2}b) drops out after the Higgs-mass
renormalization, because its loop is scale-independent.
\begin{figure}
\begin{center}
\begin{picture}(12,2.5)
\put(-2.5,-14.8){\includegraphics{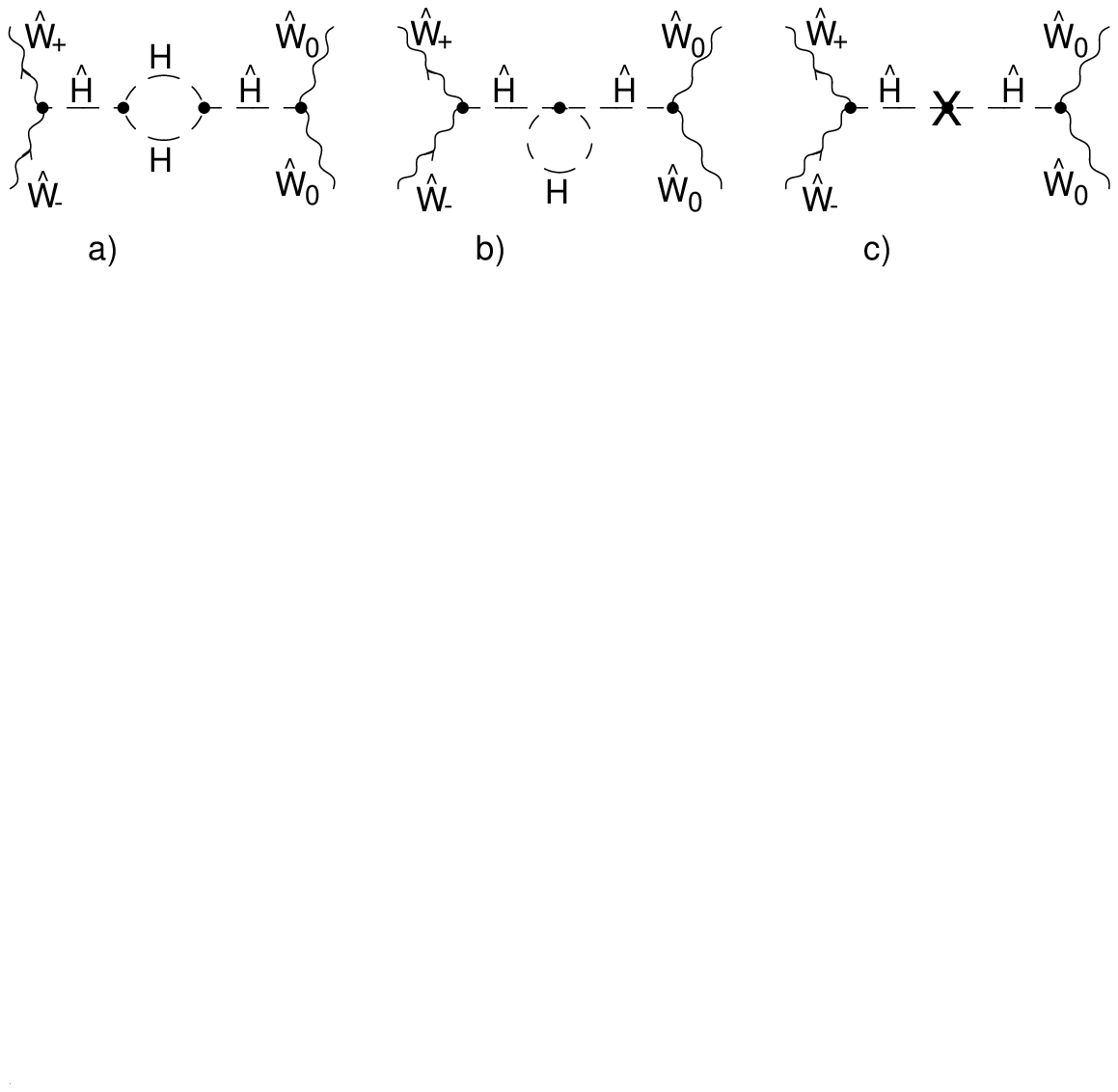}}
\end{picture}
\end{center}
\caption{Reducible Higgs diagrams for the $\hW^+\hW^-\hW^0\hW^0$
four-point-function with {\it two} background Higgs fields.}
\label{w4re2}
\efi

Summing up all contributions to
$\de\Ga^{\hW^+\hW^-\hW^0\hW^0}_{\mu\nu\rho\si}$ in $\OH{0}$, which are
given in (\ref{w4irrexp}), (\ref{w4re1exp}), (\ref{w4re2exp}), we find
the final result
\beqar
\de\Ga^{\hW^+\hW^-\hW^0\hW^0}_{\mu\nu\rho\si}
(k_+,k_-,k_{0,1},k_{0,2})
&& \nn\\
&& \hspace{-11em}
= \phantom{-} \frac{ig^4}{16\pi^2}g_{\mu\nu}g_{\rho\si}
\left[2I_{121}(\xi)+\frac{1}{2}I_{020}+2I_{222}(\xi)
+\frac{1}{\MH^2}\left(\frac{1}{4}I_{010}-I_{111}(\xi)\right)
\right. \nn\\
&& \hspace{-11em} \left.
\phantom{=-\frac{ig^4}{16\pi^2}g_{\mu\nu}g_{\rho\si}\bigg[}
-\frac{9}{8}B_0(\MH^2,\MH,\MH)
-\frac{3}{8}\Re\left\{B_0(\MH^2,0,0)\right\}\right]
\nn\\
&& \hspace{-11em} \phantom{=}
+\frac{ig^4}{16\pi^2}(g_{\mu\rho}g_{\nu\si}+g_{\mu\si}g_{\nu\rho})
\,2I_{222}(\xi).
\eeqar
This is again in agreement with the result derived from the effective
Lagrangian (\ref{leff7}). The terms in (\ref{leff7}) relevant for the
{}Feynman rules are given by
\beqar
\tr{\hW_{\mu\nu}\hW^{\mu\nu}}\Big\vert_{\hW^+\hW^-\hW^0\hW^0} &=&
2g^2(\hW^+\hW^-)(\hW^0)^2-2g^2(\hW^+\hW^0)(\hW^-\hW^0), \nn\\
\tr{\hW_{\mu\nu}[\hV^\mu,\hV^\nu]}\Big\vert_{\hW^+\hW^-\hW^0\hW^0} &=&
-2ig^3(\hW^+\hW^-)(\hW^0)^2+2ig^3(\hW^+\hW^0)(\hW^-\hW^0), \nn\\
\tr{\hV_\mu\hV_\nu}\tr{\hV^\mu\hV^\nu}\Big\vert_{\hW^+\hW^-\hW^0\hW^0} &=&
g^4(\hW^+\hW^0)(\hW^-\hW^0), \nn\\
\left(\tr{\hV_\mu\hV^\mu}\right)^2\Big\vert_{\hW^+\hW^-\hW^0\hW^0} &=&
g^4(\hW^+\hW^-)(\hW^0)^2.
\label{w4leff}
\eeqar

The given examples clearly demonstrate the advantages of the effective
Lagrangian approach for calculating the heavy-Higgs effects for specific
vertex functions of the GLSM. Instead of evaluating numerous one-loop
diagrams and expanding them for $\MH\to\infty$, one can simply read the
corresponding contributions directly from the effective Lagrangian.
In particular, it frequently happens that several types of Feynman
graphs cancel in the heavy-Higgs limit, as can be e.g.\ seen in the
previous example of the four-point function. Such contributions do not
occur in the effective Lagrangian at all.

\section{Discussion of the result}
\label{dis}
\eqnew

We find that
the limit $\MH\to\infty$ of the gauged linear $\sigma$-model
at one loop
is the gauged non-linear $\sigma$-model plus the effective
interaction terms given in \refeq{leffsmatrix}. Let us emphasize that
$\lreff$ \refeq{leffsmatrix} does {\em not} contain the complete
one-loop contributions of the GLSM at $\OH{0}$, but only those effects that
come from
diagrams
with Higgs lines. In order to find the complete
one-loop corrections to an S-matrix element within the GLSM with a
heavy Higgs boson, the contribution from the GNLSM Lagrangian
$\lag^{\rm 1-loop}_{\rm GNLSM}$ in \refeq{lfinal}, which still contains the
light quantum fields, also has to be considered.

Since the GLSM is renormalizable, all one-loop
contributions to S-Matrix
elements within this model are
UV-finite. In fact, the logartihmic
divergencies $\Delta$ (see \refeq{tde})
in \refeq{leffsmatrix} cancel against the logarithmically divergent
contributions
of the (non-re\-nor\-mal\-iz\-able) one-loop GNLSM
Lagrangian
$\lag^{\rm 1-loop}_{\rm GNLSM}$ in \refeq{lfinal}, which
have been calculated in \citere{apbe}.

In \refeq{leffsmatrix},  logarithmically  divergent contributions and
$\log\MH$-terms always occur in the
linear combination $\tde$ \refeq{tde}.
This and
the above reasoning imply that the logarithmically
divergent one-loop contributions
of the GNLSM to S-matrix elements coincide
with the logarithmically $\MH$-dependent one-loop contributions in the
GLSM,
if one replaces
\beq
\De =\frac{2}{4-D}-\gamma_E+\log(4\pi)+\log\mu^2 \quad\to\quad \log\MH^2,
\label{subs}
\eeq
as assumed in \citere{apbe}; i.e. the $\log\MH$-effects of the GLSM
can alternatively be calculated within the GNLSM. However, the
Lagrangian \refeq{leffsmatrix} contains additional finite and
$\MH$-independent contributions, which describe differences between
the GLSM and the GNLSM at one loop. Thus, the GNLSM with the
replacement \refeq{subs} is {\em not} the heavy-Higgs limit of the
GLSM at one loop, but it differs from this by finite,
constant contributions.
Since the logarithm increases very
slowly, these constants are even for a large Higgs mass
of a magnitude comparable to that of the
$\log\MH$-terms, and thus they have to be taken into account.

{}Finally, we compare our result with that found in
\citere{hemo} by diagrammatical calculations for the
electroweak standard model, which can be reduced to
the SU(2) model by setting $g'=0$ there. We find that our
result \refeq{leffsmatrix}
agrees with that of \citere{hemo}. However, it should be noted, that we
find a different factor for the
$\tr{(\DW^\mu\hV_\mu)^2}$-term in \refeq{lefffinal}.
The reason for this is that
in \citere{hemo} the linear parametrization of the
scalar sector \refeq{philin} is used,
while we applied the non-linear
parametrization \refeq{phinl}, \refeq{U}. It is well-known that such
a reparametrization of the scalar fields leaves S-matrix elements
unaffected, however, it may change Green functions \cite{stue2}. As
pointed out in the previous section, the
$\tr{(\DW^\mu\hV_\mu)^2}$-term in \refeq{lefffinal} has no effect on
S-matrix elements.
Thus, our result and that of
\citere{hemo} are consistent with each other.

\section{Summary}
\label{sum}

In this article we have described a general method to remove
non-decoupling heavy fields from a quantized field theory
at one loop
and to construct a low-energy
one-loop
effective Lagrangian by functional
methods, i.e.\ by integrating out the heavy degrees of freedom in the
path integral. We have applied this method to
a specific example, viz.\
a spontaneously broken SU(2) gauge theory, but it can immediately be
applied to any other model with a non-decoupling heavy field in order to
construct its $M\to\infty$ limit
at one loop, where $M$
is the mass of the heavy field.

We have used the background-field method, where the fields are
split into classical background fields, which correspond to tree
lines, and quantum fields, which correspond to lines inside loops. The heavy
quantum field is integrated out by performing the integration
over this degree of freedom in the path integral, while the
corresponding background field can then be removed
by a propagator expansion of its tree lines in $1/M$
or by applying the classical equations of motion in lowest order.
The resulting Lagrangian still
contains the light quantum fields, i.e.\ it does not parametrize the
complete one-loop effects of the theory but only the contribution from
loops with heavy particles. However, the effective terms
generated by integrating out the heavy field contain only background
fields, because these terms already parametrize one-loop effects, and
thus, only have to be used at tree level, when subsequently
calculating
vertex functions or
S-matrix elements at one loop.

Comparing our functional approach with diagrammatical calculations
(see \citere{hemo} and \refse{examples}), we find that it
possesses
many advantages: The
$1/M$-expansion described in \refse{helim},
i.e.\ the isolation of the
non-decoupling effects of the heavy fields from the decoupling
effects, is very easy within the functional approach. Furthermore,
our calculations could be done within the
convenient matrix notation; i.e.\
we had not to write down the components of
the fields.
This property and the application of the Stueckelberg formalism,
which removes the background Goldstone fields from intermediate
calculations, enables the simultaneous calculation of one-loop
contributions to many different Green functions.
{}For instance, our final
result also contains contributions to Green functions with external
Goldstone fields, although these fields never occurred explicitly
during
our calculation.
The use of the matrix notation and of the Stueckelberg
formalism also made it very easy to write the generated effective
Lagrangian into a manifestly gauge-invariant form.

We have applied this method to integrate out the Higgs boson
in the SU(2) gauged linear $\sigma$-model at one loop.
We have found that the
logarithmically $\MH$-dependent
contributions
to S-matrix elements
within this model coincide with the logarithmically divergent
contributions of the gauged {\it non-linear\/} $\sigma$-model if the
substitution \refeq{subs} is done, however that the latter model
differs  from the heavy-Higgs limit of the
former
by finite
and constant contributions at one loop.

As a by-product of this calculation we have formulated the
background-field method for spontaneously broken gauge theories for
the case
that the scalar sector is non-linearly parametrized, and we have
generalized the Stueckelberg formalism to the background-field
method. The renormalization has been carried out such that also the
renormalized effective action remains background-gauge-invariant.

We will apply the method described in this article to integrate out
the (heavy)
Higgs boson in the electroweak standard model in a forthcoming
article \cite{smpaper}.

\section*{Acknowledgement}
We thank Ansgar Denner and Reinhart K\"ogerler for helpful
discussions and for reading the manuscript.

\appendix
\def\theequation{\thesection.\arabic{equation}}
\setcounter{equation}{0}
\section*{Appendix}

\section{Explicit expressions for the one-loop integrals}
\label{ints}

In \refse{helim} the construction of the unrenormalized effective
Lagrangian (\ref{leff2}) was traced back to the vacuum integrals
$I_{klm}(\xi)$ defined in (\ref{intnot}). Such vacuum integrals are
easily calculated to
\beqar
&&\hspace{-2em}
\frac{(2\pi\mu)^{4-D}}{i\pi^2}\int d^D p\; \frac{\disp
p_{\mu_1}\ldots p_{\mu_{2k}}}{(p^2-M_1^2)^l}
= g_{\mu_1\ldots\mu_{2k}}\,
\frac{(-1)^{k+l}}{2^k}\,
\frac{\Ga\left(l-k-\frac{D}{2}\right)}{\Ga(l)}
      \left(4\pi\mu^2\right)^{\frac{4-D}{2}}
      M^{D+2k-2l},
\nn\\[.6em]
&&\hspace{-2em}
\frac{(2\pi\mu)^{4-D}}{i\pi^2}\int d^D p\; \frac{\disp
p_{\mu_1}\ldots p_{\mu_{2k}}}{(p^2-M_1^2)^l(p^2-M_2^2)^m}
= g_{\mu_1\ldots\mu_{2k}}\,
\frac{(-1)^k}{2^k}\,
\frac{\Ga\left(1-k-\frac{D}{2}\right)}{\Ga(l)\,\Ga(m)}
      \left(4\pi\mu^2\right)^{\frac{4-D}{2}}
\nn\\[.3em]
&& \hspace{10em} \times\,
\left(\frac{\partial}{\partial M_1^2}\right)^{l-1}
\left(\frac{\partial}{\partial M_2^2}\right)^{m-1}
\left[\frac{M_1^{D+2k-2}-M_2^{D+2k-2}}{M_2^2-M_1^2}\right],
\hspace{2em}
\label{vacint}
\eeqar
for arbitrary space-time dimension $D$.
According to (\ref{vacint}) the relevant
$\OH{0}$ parts of the $I_{klm}$ for $D\to 4$
are given by
\beqar
I_{010}\phantom{(\xi)}&=&\MH^2(\tde+1),\nn\\
I_{011}(\xi)&=&\tde+1 +\OH{-2},\nn\\
I_{020}\phantom{(\xi)}&=&\tde,\nn\\
I_{111}(\xi)&=&\frac{1}{4}(\MH^2+\xi\MW^2)
             \left(\tde+\frac{3}{2}\right)+\OH{-2},\nn\\
I_{112}(\xi)&=&\frac{1}{4}\left(\tde+\frac{3}{2}\right)+\OH{-2},\nn\\
I_{121}(\xi)&=&\frac{1}{4}\left(\tde+\frac{1}{2}\right)+\OH{-2},\nn\\
I_{213}(\xi)&=&\frac{1}{24}\left(\tde+\frac{11}{6}\right)+\OH{-2},\nn\\
I_{222}(\xi)&=&\frac{1}{24}\left(\tde+\frac{5}{6}\right)+\OH{-2},
\label{integrals}
\eeqar
with
\beq
\tde=\De-\log\left(\frac{\MH^2}{\mu^2}\right), \qquad
\De =\frac{2}{4-D}-\gamma_E+\log(4\pi),
\label{tde}\eeq
and $\gamma_E$ being Euler's constant.

In \refse{ren} we expressed the renormalization constant $\de\MH^2$
(\ref{dmh}) in terms of $I_{klm}$ and scalar two-point functions
$B_0(k^2,M_1,M_2)$ defined in (\ref{b0}). The explicit expressions for
the relevant $B_0$-functions can for instance be deduced from the general
result presented in \citere{de93}, leading to
\beqar
B_0(\MH^2,\MH,\MH) &=& \tde+2-\frac{\pi}{\sqrt{3}},
\nn\\
B_0(\MH^2,0,0)     &=& \tde+2+i\pi.
\label{Bs}\eeqar


\begin{thebibliography}{00}
\frenchspacing

\bibitem{hemo}M.J.~Herrero and E.~Ruiz Morales, \nphb{418} (1994)
431; Madrid preprint FTUAM 94/11 (1994), hep-ph/9411207

\bibitem{bisa}M.~Bilenky and A.~Santamaria, \nphb{420} (1994) 47

\bibitem{gale}J.~Gasser and H.~Leutwyler, \aph{158} (1984) 142;\\
A.~Nyffeler and A. Schenk, Bern Preprint BUTP-94/12 (1994),
hep-ph/9409436

\bibitem{fume}I.J.R.~Aitchison and C.M.~Fraser, \phlb{146} (1984) 63;
\phrd{31} (1985) 2605;\\
C.M.~Fraser, \zphc{28} (1985) 101;\\
J.A.~Zuk, \phrd{32} (1985) 2653; {\bf D33} (1986) 3545;\\
O.~Cheyette, \phrl{55} (1985) 2394;\\
M.K.~Gaillard, \nphb{268} (1986) 669;\\
L.-H.~Chan, \phrd{36} (1987) 3755;\\
R.D.~Ball, \phrp{182} (1989) 1

\bibitem{chan}L.-H.~Chan, \phrl{54} (1985) 1222
[Err.\ {\bf 56} (1986) 404];
{\bf 57} (1986) 1199

\bibitem{chey} O.~Cheyette, \nphb{297} (1988) 183

\bibitem{bfm1}B.S. DeWitt, \phr{162} (1967) 1195; {\em Dynamical
Theory of groups and Fields\/} (Gordon and Breach, New York, 1965);
in {\em Quantum Gravity 2}, ed. C.J.~Isham, et. al.\ (Oxford
University Press, New York, 1981), p.~449;\\
G.~'t Hooft, Acta Universitatis Wratislavensis {\bf 368} (1976) 345;\\
H.~Kluberg-Stern and J.~Zuber, \phrd{12} (1975) 482 and 3159;\\
D.G.~Bouleware, \phrd{23} (1981) 389;\\
C.F.~Hart, \phrd{28} (1983) 1993

\bibitem{bfm2}L.F.~Abbott, \nphb{185} (1981) 189; Acta.\ Phys.\ Pol.\
{\bf B13} (1982) 33;\\
L.F.~Abbott, M.T.~Grisaru and R.K.~Schaefer, \nphb{229} (1983) 372

\bibitem{bfm3}M.B.~Einhorn and J.~Wudka, \phrd{39} (1989) 2758

\bibitem{bfm4}A.~Denner, S.~Dittmaier and G.~Weiglein, \phlb{333}
(1994) 420;
\nphbps{37} (1994) 87

\bibitem{bfm5}A.~Denner, S.~Dittmaier and G.~Weiglein,
Bielefeld Preprint 94/50 (1994), hep-ph/9410338

\bibitem{smpaper}S.~Dittmaier and C.~Grosse-Knetter, in preparation

\bibitem{bash}W.A. Bardeen and K. Shizuya, \phrd{18} (1978) 1969

\bibitem{apbe}T. Appelquist and C. Bernard,
Phys.\ Rev.\ {\bf D22} (1980) 200;\\
A. C. Longhitano, Nucl.\ Phys.\ {\bf B188} (1981) 118

\bibitem{stue1}E. C. G. Stueckelberg,
Helv.\ Phys.\ Acta {\bf 11} (1938) 299; {\bf 30} (1956) 209;\\
T. Kunimasa and T. Goto, Prog.\ Theor.\ Phys.\ {\bf 37} (1967) 425

\bibitem{stue2}B. W. Lee and J. Zinn-Justin,
Phys.\ Rev.\ {\bf D5} (1972) 3155;\\
C. Grosse-Knetter and R. K\"ogerler,
Phys.\ Rev.\ {\bf D48} (1993) 2865

\bibitem{ao80}
K.I.\ Aoki, Z.\ Hioki, R.\ Kawabe, M.\ Konuma and T.\ Muta,
\ptph{64} (1980) 707; {\bf 65} (1981) 1001;
Suppl.\ \ptph{73} (1982) 1;\\
M.~B\"ohm, W.~Hollik and H.~Spiesberger, \fp{34} (1986) 687

\bibitem{de93}
A.\ Denner, \fp{41} (1993) 307

\bibitem{eom}D. Barua and S. N. Gupta, Phys.\ Rev.\ {\bf D16} (1977) 413;\\
H. D. Politzer, Nucl.\ Phys.\ {\bf B172} (1980) 349;\\
H. Georgi, Nucl.\ Phys.\ {\bf B361} (1991) 339;\\
C. Arzt, Michigan-Preprint UM-TH-92-28 (1992), hep-ph/9304230;\\
H. Simma, Z. Phys.\ {\bf C61} (1994) 67;\\
C.~Grosse-Knetter, \phrd{49} (1994) 1988 and 6709

\end{thebibliography}
\end{document}